\documentclass[preprint,12pt,authoryear]{elsarticle}

\usepackage{amssymb}
\usepackage{amsmath}

\usepackage[dvipsnames,table]{xcolor}
\usepackage{colortbl} 
\usepackage{hyperref}
\usepackage{url}

\usepackage[normalem]{ulem}
\usepackage{arydshln}
\usepackage{lmodern}
\usepackage{tabularx,booktabs,ragged2e}
\usepackage{makecell}
\usepackage{multirow}
\newcolumntype{C}{>{\centering\let\newline\\\arraybackslash\hspace{0pt}}m{2cm}}
\usepackage{enumitem}
\DeclareUnicodeCharacter{221A}{\textendash}
\journal{Land Use Policy}

\begin{document}

\begin{frontmatter}



\title{Modern approaches to building interpretable models of the property market using machine learning on the base of mass cadastral valuation} 


\author[bimsa]{Tanashkin Alexey S.\corref{cor1}}
\ead{tanashkin.alexey@gmail.com}
\cortext[cor1]{Corresponding author}
\author[self]{Tanashkina Irina G.} 
\author[ccv]{Maksimchuik Alexander S.}

\affiliation[bimsa]{organization={Beijing Institute of Mathematical Sciences and Applications},
            addressline={No. 544, Hefangkou Village Huaibei Town}, 
            city={Huairou District Beijing},
            postcode={101408}, 
            state={Beijing},
            country={China}}
\affiliation[self]{Independent Researcher}
\affiliation[ccv]{organization={Center of cadastral valuation for Primorsky Krai},
            addressline={Ostryakova prospekt 49, room 505}, 
            city={Vladivostok},
            postcode={690078}, 
            state={Primorsky Krai},
            country={Russia}}

\begin{abstract}
In this paper, we review modern approaches to building interpretable models of property markets using machine learning on the base of mass valuation of property in the Primorye region, Russia. There are numerous potential difficulties one could  encounter in the effort to build a good model. Their main source is the huge difference between noisy real market data and ideal data usually used in tutorials on machine learning. This paper covers all stages of modeling: collection of initial data, identification of outliers, search and analysis of patterns in the data, formation and final choice of price factors, building of the model, and evaluation of its efficiency. For each stage, we highlight potential issues and describe sound methods for overcoming emerging difficulties on actual examples. We show that the combination of classical linear regression with kriging (interpolation method of geostatistics) allows to build an effective model for land parcels. For flats, when many objects are attributed to one spatial point, the application of geostatistical methods becomes problematic. Instead, we suggest linear regression with automatic generation and selection of additional rules on the base of decision trees, so called the RuleFit method. We compare the performance of our inherently interpretable models with well-proven ``black-box'' Random Forest method and demonstrate similar results. Thus we show, that despite such a strong restriction as the requirement of interpretability which is important in practical aspects, for example, legal matters, it is still possible to build effective models of real property markets.
\end{abstract}



\end{frontmatter}



\section{Introduction}
\label{intro}

The problem of estimating the real estate value has existed for a long time and is actual nowadays. It can be approached differently and include different constraints that impact predicted values. Among them are the aim to get the most accurate predictions; the limit on the algorithm runtime; the interpretability of the model, and so on. The prediction process is additionally complicated by the insufficiency of data, underdevelopment of the market, lack of explanatory factors for prediction of the real estate value, and so on. There are also problems with the nonlinear influence of some variables on others and low correlation between explanatory variables and the dependent variable~\citep{Eren2023}.

The scope of research may vary from forecasting time series \citep[e.g. ][]{Kouwenberg2014,Wei2017,Milunovich2020} and  comprehensive analysis of evolution trends of real estate markets taking into account various economical factors
\citep{Song2025}, up to mass appraisal of real estate \citep[see ][]{Kontrimas2011,McCluskey2013, Arribas2016,Wang2019}.
The latter can be applied for inspection of the property market or administrative regulations such as property taxes. In most cases in the Russian Federation property taxes are determined from the cadastral value of the object which is calculated in the process of cadastral valuation. Cadastral valuation allows to create a clear and valid system of taxation and other relations governed by law.
It significantly influences the economic development of territories in municipalities. 
Incorrect cadastral valuation can slow down the development of particular areas and create financial and economic disbalance~\citep{Byrda2025}.

Cadastral valuation is conducted by methods verging its predictions to market values as close as possible. The market value of an object can vary significantly over time. Due to this reason, the cadastral value of an object should be regularly updated through mass cadastral valuation of real estate in order for the taxes to be accurate and fair \citep{Jahanshiri2011}. The result of such mass appraisal is the model based on market data, which predicts the price of the arbitrary object. In this paper, we discuss methods of effective mass valuation of real estate on the example of the state cadastral valuation of property in the Primorye region, Russia, which is regulated by certain methodological rules.
According to them, the prediction of the real estate value should be made using one of the following methods: the comparative method, the cost method, or the income method. The comparative method is typically used in the case when the market of real estate property is highly developed and there is enough information about deals and offers. This method includes the method of a fiducial object and regression analysis tools such as the linear regression and its derivatives~\citep{Method2018}. This generally agrees with the guidelines of \cite{IAAO2025}, although many authors tend to place linear models into a separate category \citep{Pagourtzi2003,Deppner2023}. Since \cite{Rosen1974} introduced hedonic pricing approach to the value of an object as the sum of the implicit prices of its objectively measurable characteristics, linear models became a powerful tool for assessing the real estate market~\citep{Adair1996}. However, the linear nature of models leads to their inability of capturing non-linear relationships~\citep[e.g. ][]{Mark1988,Dunse1998,Limsombunc2004,Zurada2011}.  In order to address this problem, for the last several decades researchers turn their attention to advanced machine learning\footnote{While the linear regression was well established long before machine learning was introduced, nowadays it is often referred as ML model \citep[e.g. ][]{Molnar2022}.} (ML) methods.

There is a vast amount of research on the application of machine learning to prediction of property values \citep{MorenoForonda2025}. Among the most established ML models are the Support Vector Machine \citep[SVM, ][]{Smola2004,Lam2009,Li2009,Gu2011,Phan2018}, the Random Forest \citep[RF, ][]{Breiman2001,Antipov2012,Ho2020,Hong2020,Levantesi2020,Shokoohyar2020,Sweta2023}, the Extreme Gradient Boosting \citep[XGBoost or XGB, ][]{Chen2016,Sing2021,Joseph2024,Yarmatov2024,Sharma2024,Kee2025}, and Gaussian Process Regression \citep[GPR, ][]{Rasmussen2005,Crosby2016,Xu2023,Dearmon2024,Jin2025}. Additionally, the Artificial Neural Networks (ANN) constitute the separate extensive class of models and have been found to perform well in housing prices prediction ~\citep{Zurada2011,Forys2022,Xu2022,Sweta2023}. Comparative studies generally report the advantage of the mentioned models over the linear regression \citep{Yilmazer2020, Kalaivani2023, Lahmiri2023, Cui2024, Soegianto2024, Dawid2025}. However, despite the impressive predicting ability, the implementation of advanced machine learning techniques for property policies remains rather limited due to the ``black-box'' nature of the ML models \citep{DoshiVelez2017}. The term ``black-box'' was coined by \cite{Ashby1956} in scope of Cybernetics to describe the task of investigating the unknown system by inferring how different inputs influence the output, and it perfectly reflects the opaque mechanisms of model predictions. 
At the same time, if the model forms a base for legal policies (e.g. taxing), then transparency becomes its key aspect. In this context, transparency means interpretability.
After the high potential of ML in terms of catching the complex patterns in real estate data was realized, the topic of interpretability has gained serious attention, and some common model-agnostic post-hoc methods have been established. They are characterized by universality and can be applied to a broad range of models. These methods utilize the SIPA principle: Sample from the data, perform an Intervention on the data, get the Predictions for the manipulated data, and Aggregate the results \citep{Scholbeck2020}. This is a canonical example of application of the ``black-box'' concept: analyzing how varying of input data changes the output, one can elucidate the inner mechanics of a model.

Depending on the goal, model-agnostic methods are divided into global and local. The former describe the average behavior of a model on the whole dataset allowing to investigate relationships between features and target variable (feature effects), and to rank features by their importance (feature importance). Recent papers \citep{Lorenz2022, Deppner2023, Mathotaarachchi2024,Song2025,Kee2025} show the superb ability of these methods to describe the property markets, to highlight the most significant factors influencing prices formation, and to reveal feature interactions. So the global methods are a powerful tool for policymakers and investors to capture large-scale trends, offering scientific basis for market regulation and investment decisions \citep{Hu2019, Song2025}. Nevertheless, one should be aware of the limitation of this approach -- it only allows to explain the model behavior, which can differ from the real market situation. 

In contrast to global methods, local model-agnostic methods serve to explain the price formation of a specific property object. This task is superior in the context of the cadastral valuation, where the price determinants should be transparent to property owners.  
The examples of local methods are the Ceteris Paribus Plots \citep{Kuzba2019}, Individual Conditional Expectation \citep{Goldstein2015}, and Counterfactual Explanations \citep{Wachter2017,Dandl2020}. The detailed description of them, as well as the global methods, can be found in the \cite{Molnar2022} handbook. However, the mentioned techniques are more suitable for inferring the influence of one particular feature on the target variable.
For getting the comprehensive representation of how all features contribute to the price, the Local Interpretable Model-agnostic Explanations (LIME) and the SHapley Additive exPlanations (SHAP) are used. The LIME introduced in \cite{Ribeiro2016} belongs to the family of local surrogate models. It generates a new sample by perturbing the data in the vicinity of the point of interest (the particular value to be explained) and evaluates the model on this new data set. Next, the inherently interpretable model (e.g. linear regression or decision tree) is trained on the resulting data and is used to interpret this specific point. This is a promising approach, but on the current state of development it has serious limitations. The problem of defining the vicinity of the point has not been solved yet, and as it was shown by \cite{AlvarezMelis2018} the explanations of two neighboring points can be very different. Additionally, as was pointed out by \cite{Molnar2022}, repeating of sampling process can lead to different results. Finally, contributions of features of the surrogate model do not sum up to the value predicted by the initial ``black-box'' model. All of this makes the implementation of LIME to cadastral valuation, where the evaluation directly serves for tax calculation, questionable, at least. In this context, the SHAP \citep{Lundberg2017} seems to have better perspectives. This method is based on the rigorous concept of Shapley values, which were initially suggested as an approach to fair distribution of payout among the players (according to their contribution to a total payout) in game theory \citep{Shapley1953}. Later, they were introduced by \cite{Strumbelj10a} as an instrument for interpreting ML predictions. Originally the Shapley value for a player (a feature value) is calculated as the weighted average of the marginal contribution of the feature value to all possible coalitions -- sets of features included in the model, starting from no features, only one feature, all possible combinations of two features, etc. The marginal contribution of a feature to a coalition (a set of features) is the difference between the predictions of the model trained on the set of features including this particular one and without it. 
Among other properties, Shapley values obey the property of efficiency, which means that the sum of Shapley values for all features is equal to the difference between the prediction of original ``black-box'' model for the particular instance and the average of model predictions over the whole dataset. Therefore, each Shapley value shows the contribution of the corresponding feature to the deviation of the predicted price from the average value, making the prediction of the original ``black-box'' model completely interpretable. However, in practice, the situation is more complicated. Calculating the Shapley values requires retraining the model for assessing the marginal contribution for all possible coalitions of features, and this number grows exponentially with the total number of features included in the model, quickly making the direct computation unfeasible. Recently, some model-specific algorithms for exact computation of Shapley values in polynomial time were presented \citep{Mastropietro2023, Mohammadi2025}. For model-agnostic solutions, various approximation methods are suggested \citep{Qin2025}. To avoid retraining the model, the absent features in a coalition are replaced by corresponding feature values randomly sampled from the dataset. Next, sophisticated sampling from the set of all possible coalitions is implemented. Currently, there are three main methods -- the originally proposed KernelSHAP \citep{Lundberg2017}, which used LIME for approximating Shapley values, and, despite being still quite slow, encouraged wide implementation of Shapley values in machine learning; the TreeSHAP method \cite{Lundberg2018}, which is fast but works only with tree-like models such as Random Forest or XGBoost, so it is model specific; and Permutation method \cite{Mitchell2022}, which takes advantage of equivalent definition of Shapley values in terms of permutations of features and efficiently samples from a set of all permutations. Given all arguments above, the SHAP is a powerful tool for sound explanations of ``black-box'' models, but its usage to mass cadastral valuation is still not obvious due to complexity of the method. While it provides the property owner with the information on how each feature contributes to the total model prediction, these arguments may not withstand if the owner decides to challenge the assigned value in court. Legal aspects of machine learning are not well established yet, so the judge may not take this evidence into account, preferring traditional methods like the appraisal report with the evaluation of the complainant property conducted using, for example, the comparative method.

Given the aforementioned challenges, in this paper we take a different approach and instead of post-hoc interpretation methods consider the models interpretable by design, also known as the inherently interpretable models. Examples of such models are the linear regression and its extensions with addition of penalties, interactions and non-linear terms; or the decision trees, which are basic components of the Random Forest. Design of these models allows simple interpretation and does not require expertise and deep understanding of complex methods, fostering their wide implementation in various applied tasks demanding rigorous but simple justification. However, the cost of simplicity is generally inferior performance comparing to more advanced algorithms. As underlined in \cite{Jahanshiri2011} review of mass valuation models, one of the major limitations of linear models is poor treatment of spatial autocorrelations and heterogeneity of data, which is crucial in the case of real estate property. For overcoming this issue, the spatial extensions of the linear regression are used. Spatial AutoRegressive (SAR) models \citep{Dubin1999} tackle spatial autocorrelations by adding to the OLS equation the additional term containing the spatial weight matrix, resulting in more precise predictions \citep{Kobylinska2021}. However, it complicates interpretation of the results. Another firmly established approach is the Geographically Weighted Regression ~\citep[GWR, ][]{Brunsdon1996, Sisman2022} dealing with the spatial heterogeneity by splitting the data in local regions (so called ``windows'') and building a linear regression model for each window weighting the observations according to their distance to the regression point. This procedure leads to a number of regression models each having a different set of coefficients, which also complicates the model. Another issue is that there are various methods of splitting the data in regression windows, potentially leading to different results. SAR and GWR can be successfully combined \citep{Tomal2020}. Another promising possibility is to incorporate these spatial methods into ML algorithms such as Random Forest \citep{Credit2021, Georganos2019, Santos2019} or Neural Networks  \citep{Wang2022}. Unfortunately, interpretability is not a strong point of these methods.

In order to find an appropriate trade-off between intrinsical interpretability and precision, we use the ordinary least squares linear regression to build a base model. It allows to highlight the general trend in the data and explain how independent features influence predictions. Next, we model the residuals by kriging \citep{ChicaOlmo2007,Demidova2013}, which is a special case of the Gaussian Process Regression \citep{Rasmussen2005} used for spatial interpolation in Geostatistics~\citep{Cressie2015}. This combined approach is known as the regression-kriging~\citep{Hengl2007,ChicaOlmo2020}, and it allows to keep the interpretability of the model on the one hand and increase its precision on the other. We show that the implementation of this method for modeling of land parcels leads to improvement of predictions comparing to the linear model.

Besides land parcels, we also consider another large market segment -- flats in residential buildings. In this case, combining the linear regression and spatial methods is challenging since normally many flats correspond to one spatial point that is the location of a residential building. Instead, we adapt the RuleFit method \citep{Friedman2008}, which combines the Random Forest and the linear regression with additional penalty for complexity of the model. The Random Forest is used to generate combinations of features (many of them describe spatial attributes of the data in terms of distance to something) to create a set of rules. These rules along with the initial features are used to train the sparse linear model, which finally contains the set of the most important features. Such approach allows to take into account the spatial effects and feature interactions and also, to some extent, to address non-linear relationships in the data. To the best of our knowledge, the current study is the first implementation of the RuleFit to the real estate segment.

\begin{table}[!t]
\begin{center}
\caption{Overview of the methodology implemented in the research. Enumerated lists reflect the consecutive steps while itemized ones do not require strict order.}
\label{tab:tab_sum}
\scriptsize
\begin{tabular}{ |p{0.7em}|m{0.5em}|m{8.5em}|m{15em}|m{15em}| }
\hline
\multicolumn{3}{|c|}{\textbf{Steps}} & \textbf{Land parcels} & \textbf{Flats} \\
\hline
\textbf{I} &\multicolumn{2}{m{9em}|}{Collection of data} & \multicolumn{2}{m{30em}|}
{
\begin{minipage}[t]{\linewidth}
\vspace{1pt}
\begin{itemize}[noitemsep,leftmargin=*,after=\strut]
    \item Offers: collected by a third-party company by parsing internet platforms
    \item Deals: provided by Ministry of Property and Land Relations of Russia
\end{itemize}
\end{minipage}
} \\
\hline
\textbf{II} & \multicolumn{2}{m{9em}|}{Detection of outliers} &
\begin{minipage}[t]{\linewidth}
\begin{enumerate}[nosep,leftmargin=*,after=\strut]
    \item Spatial clustering (k-means)
    \item Standard statistical methods (box-plot, z-score)
\end{enumerate}
\end{minipage}
& 
\begin{minipage}[t]{\linewidth}
\begin{enumerate}[nosep,leftmargin=*,after=\strut]
    \item Analysis of PDF of the target variable
    \item Comparison of objects inside one building
    \item Spatial clustering (k-means)
    \item Clustering in the feature space (DBSCAN)
    \item RANSAC regression
\end{enumerate}
\end{minipage}
\\\hline

\multirow{2}{0.7em}{
\begin{minipage}[c]{\linewidth}
\vspace{5.8em}
\begin{center}
\textbf{III}
\end{center}
\end{minipage}
} & \multirow{2}{0.5em}{
\begin{minipage}[t]{\linewidth}
\vspace{2em}
\rotatebox[origin=l]{90}{\scriptsize{Feature engineering}}
\end{minipage}
}
&  \multicolumn{1}{p{8.5em}|}{
\begin{minipage}[t]{\linewidth}
\vspace{3em}
Formation of new features
\end{minipage}
} &
\multicolumn{1}{p{15em}|}{
\begin{minipage}[t]{\linewidth}
\vspace{2.5em}
\begin{itemize}[nosep,leftmargin=*,after=\strut]
    \item Aggregation of simple features
    \item Principal component analysis
\end{itemize}
\end{minipage}
}
& 
\multicolumn{1}{p{15em}|}{
\begin{minipage}[t]{\linewidth}
\vspace{1pt}
\begin{itemize}[nosep,leftmargin=*,after=\strut]
    \item Aggregation of simple features
    \item Principal component analysis
    \item Usage of OSM data for spatial features
    \item Usage of methods from related fields (graph theory, geostatistics)
\end{itemize}
\end{minipage}
}\\\cline{3-5}
& & \multicolumn{1}{p{8.5em}|}{
\begin{minipage}[t]{\linewidth}
\vspace{2em}
Selection of features
\end{minipage}
} & \multicolumn{2}{p{30em}|}
{
\begin{minipage}[t]{\linewidth}
\vspace{0.5pt}
\begin{itemize}[nosep,leftmargin=*,after=\strut]
    \item Analysis of the correlation matrix (removing multicollinearity, selection of the most significant features, removing features which complicate the interpretation)
    \item Analysis of the target variable PDFs for particular features
    \item SelectKBest algorithm
    \item Recursive Feature Elimination
\end{itemize}
\end{minipage}
} \\\hline
\textbf{IV} & \multicolumn{2}{m{9em}|}{Modeling} &
\begin{minipage}[t]{\linewidth}
\begin{itemize}[nosep,leftmargin=*,after=\strut]
    \item OLS regression
    \item OLS regression + kriging
    \item Random Forest (for comparison)
\end{itemize}
\end{minipage}
& 
\begin{minipage}[t]{\linewidth}
\vspace{1pt}
\begin{itemize}[nosep,leftmargin=*,after=\strut]
    \item OLS regression
    \item RuleFit (Random Forest for generation of feature combinations + LASSO regression)
    \item Random Forest (for comparison)
\end{itemize}
\end{minipage}
\\
\hline
\begin{minipage}[t]{\linewidth}

\vspace{2.8em}

\textbf{V} 
\end{minipage}
& 
\multicolumn{2}{p{9em}|}{
\begin{minipage}[t]{\linewidth}
\vspace{2.8em}
Evaluation of results
\end{minipage}
}
& \multicolumn{2}{p{30em}|}
{
\begin{minipage}[t]{\linewidth}
\vspace{0.1em}
\begin{itemize}[nosep,leftmargin=*]
    \item Splitting data in test and train sets, cross-validation
    \item Checking realization of model assumptions
    \item Mean Absolute Error (MAE)
    \item Mean Absolute Percentage Error (MAPE)
    \item Coefficient of determination $R^2_{\mathrm{adj}}$
    \item Comparison of interpretable models with a "black box" type model
\end{itemize}
\vspace{0.3em}
\end{minipage}
}
\\
\hline
\end{tabular}
\end{center}
\end{table}

As was noted by \cite{Mathotaarachchi2024}, the existing literature lacks of the cohesive presentation of the process of the ML-based real estate price prediction, from general methodologies and particular methods to treatment of the real, imperfect data. We share this point of view and in this paper systematically review all stages of the modeling process, highlight the main problems the appraiser can encounter, and offer sound solutions with special focus on data preparation. We present a new approach to identifying outliers in the highly noisy data, which is demonstrated on the base of flats. Additionally, we propose ways of constructing the new features and combining the existing ones. They are useful when the initial data contain a small number of features or the features have low correlation with the target variable. Among other features, we suggest how to define the ``Development of the road network'' feature on the base of the concept of centrality from graph theory. Finally, we compare the performance of our inherently interpretable models and the Random Forest. The methodology of our research is summarized in Table~\ref{tab:tab_sum}. The paper is organized as follows. In Section~\ref{res_target} we briefly introduce the territory under consideration and shortly describe the data for modeling. Our approach of handling outliers is presented in Section~\ref{outliers}. Section~\ref{select_variables} focuses on details of feature engineering. In Section~\ref{modeling} we present the results of our modeling for the segment ``Land parcels'' and ``Flats'' and discuss arising difficulties. We make the conclusion in Section~\ref{conc}.

\section{Description of the target of research}
\label{res_target}

\begin{figure}[t]
    \includegraphics[width=\linewidth]{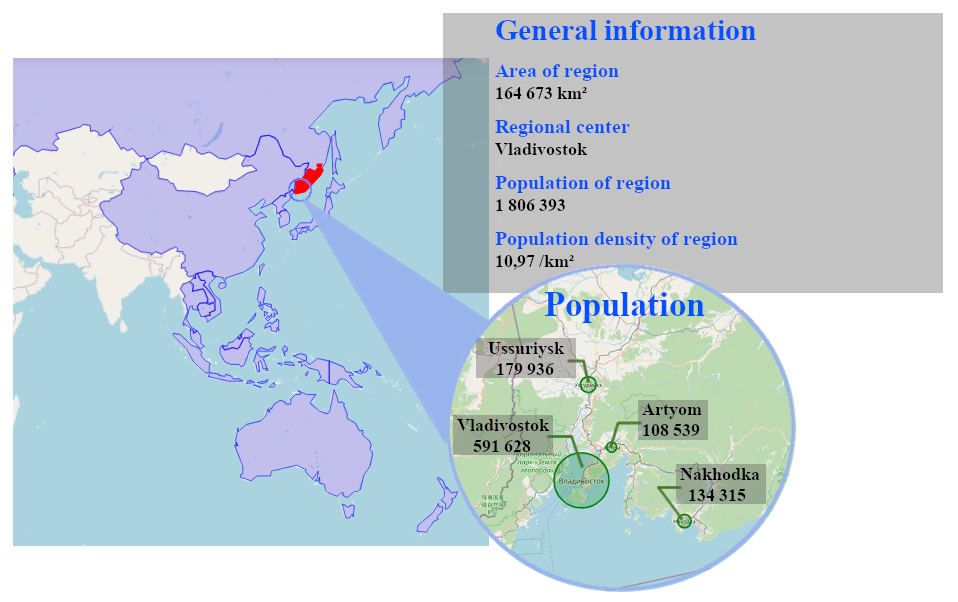}
    \caption{Primorsky Krai (highlighted in red) on the map of the Asia-Pacific region. Countries geographically located in this region are highlighted in purple. Map lines do not necessarily depict accepted national boundaries.}
    \label{fig:about_prymorye}
\end{figure}

The modeling territorially covers Primorsky Krai –- the administrative region of the Russian Federation located in the southeast of the country (Figure~\ref{fig:about_prymorye}). The region borders China on the west, North Korea on the south-west, and Khabarovsk Krai (Russian Federation) on the north. Primorsky Krai is the part of the Far Eastern Federal District. The area of the region is 164 673 km$^2$ which is comparable to areas of such countries as Bulgaria (110 994 km$^2$), Greece (131 957 km$^2$) or Cambodia (181 035 km$^2$). The population of the region is \(\sim\) 1.8 million, its administrative capital is the port city Vladivostok with population of \(\sim\) 590 thousand. Vladivostok is the political, cultural, scientific, and educational center of the region. In 2012 Vladivostok was a venue for the Asia-Pacific Economic Cooperation (APEC) summit. For this event a wide range of significant measures was taken –- development of telecommunications, construction of new roads, enhancement of the cultural and prestigious status of the region (such as building of the Opera and Ballet Theater, building of multipurpose health center, etc.), reconstruction of energy and heat supply facilities contributing to the reduction of energy costs of enterprises and energy losses in public utilities. Vladivostok is one of ten cities in the country where the Federal University was created (its new campus on the Russky Island was the venue for the APEC). This university has become a powerful intellectual base for the region.

The membership of Russia in the APEC, and Primorsky Krai in particular as the main territory for interaction with other countries of the Asia-Pacific region, allows balanced management of the regional economy and effective use of the natural resource potential of the Far East of Russia. This in turn increases the attractiveness of the region as a place for residence and creates opportunities for enhancement of the business availability and for attraction of potential investors~\citep{Chebotarev2012}. Additionally, the APEC favors regional economic integration, supports a multilateral trading system, unlocks the potential of the digital economy, and helps to bridge the digital divides~\citep{ApecURL}.

All of the aforementioned indicate the development of Primorsky Krai and the economic growth of the region. To facilitate this process each sector of economics should work effectively. Because of this reason, the results of cadastral valuation must show the real market situation, create a valid taxation system, and be a base for other relations governed by law.

Precisely because of its favorable territorial location, Vladivostok being in the third ten of Russia cities by population and separated by thousands of kilometers from the central part of Russia, stably occupies a place in the top ten of cities of Russia by real property value. But precise modeling of real property value is a difficult task due to several factors: complex terrain, historical-geographical urban development, non-uniform transport accessibility of different parts of the city, existence of insular territories, mixing of historic and modern buildings, non-uniform density of social, cultural and entertainment centers, coexistence of industrial estate and domestic development, and so on. The requirement of interpretability of the model additionally complicates the task. Thus, at each stage one should use a wide set of machine learning methods to build an effective model of property values for Vladivostok.

Machine learning methods help to find the patterns and rules and to extend them to the general sample. They are applicable in the case of sufficient amount of data. Because of this reason, we consider only two mass market segments for building the models -- flats and land parcels for private housing projects. The dependent variable for prediction is the real estate value, specifically the Per Square Meter Price (PSMP) of real estate:

\begin{equation}
    \mathrm{PSMP} = \text{total real estate value}/\text{area of real estate property}.
    \label{eq:psmp}
\end{equation}

In practice, analysts prefer to model the PSMP and not the total estate price. The reason is that the area of the property is a fairly significant pricing variable and non-inclusion of it in the model often results in the decrease of quality of the model. But if one considers the total real estate price and simultaneously includes the area into a list of independent variables then it completely suppresses other variables with no contribution to the predictive power of the model.

Basic data for the real estate value were taken from open sources (like websites with offers for the sale of property) and deals. Information about deals was presented by the Ministry of Property and Land Relations of Russia according to the federal law. We used the data collected for one year before the modeling. Thus we formed two samples for modeling –-  26432 objects for the segment “Flats” and 19246 objects for the segment “Land parcels”. For each segment, we created and analyzed explanatory variables and their correlation relationships,  searched for patterns, and built models.

\section{Identification of outliers}
\label{outliers}

\begin{figure}[t]
    \includegraphics[width=\linewidth]{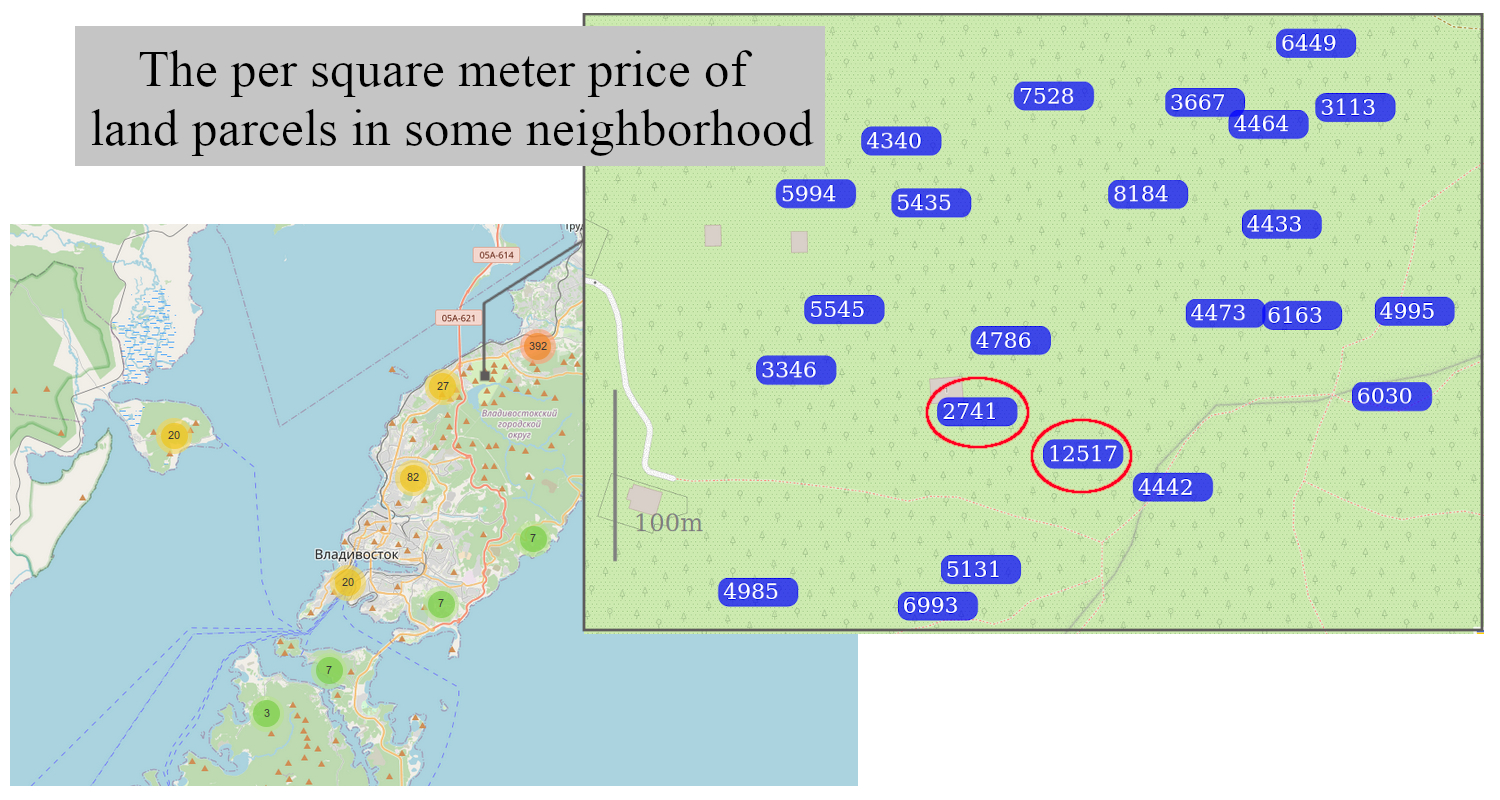}
    \caption{Fluctuation of the market per square meter price (in RUB) for the segment “Land parcels”. Significantly different prices of adjacent land parcels are marked by red ovals to guide the eye. The values in filled circles show the number of objects for a particular area. The triangular symbols represent hills and are added to the map automatically as part of the standard OpenStreetMap~\citep{OSM} layout. }
    \label{fig:fig2}
\end{figure}

An important step in modeling is checking data for validity. The data are prone to mistakes of several origins.  The most common of them are accidental mistakes (typos and so on) and intentional distortion of data by publishing fictitious values –- too high or too low. As a result, the process of creating a model becomes more difficult and may lead to incorrect results. It is worth noting that in the process of analyzing the data on the outliers, one should take into account the type of objects. A land parcel loosely can be considered as an one-dimensional object. 
We mean that basically the geographic coordinates of a land parcel define its price. Coordinates decode both the spatial location of the object and its individual characteristics such as the slope, the availability of electricity, and so on. There is one to one mapping between coordinates and the land parcel. On the other hand, the typical multi-unit residential building contains several tens or even hundreds of flats which means that one spatial point corresponds to multiple objects. Flats in the same building can differ by the storey or the area. Even the adjacent buildings having almost the same coordinates can differ drastically by the year of built, the wall material, etc. This is fairly common situation for flats and rather untypical for land parcels where the neighboring objects tend to have similar characteristics. As a result, the process of identification of outliers for the former can be more challenging. For the latter case, we can plot the per square meter price of land parcels on a map and estimate its validity from general considerations.

\begin{figure}[t]
    \includegraphics[width=\linewidth]{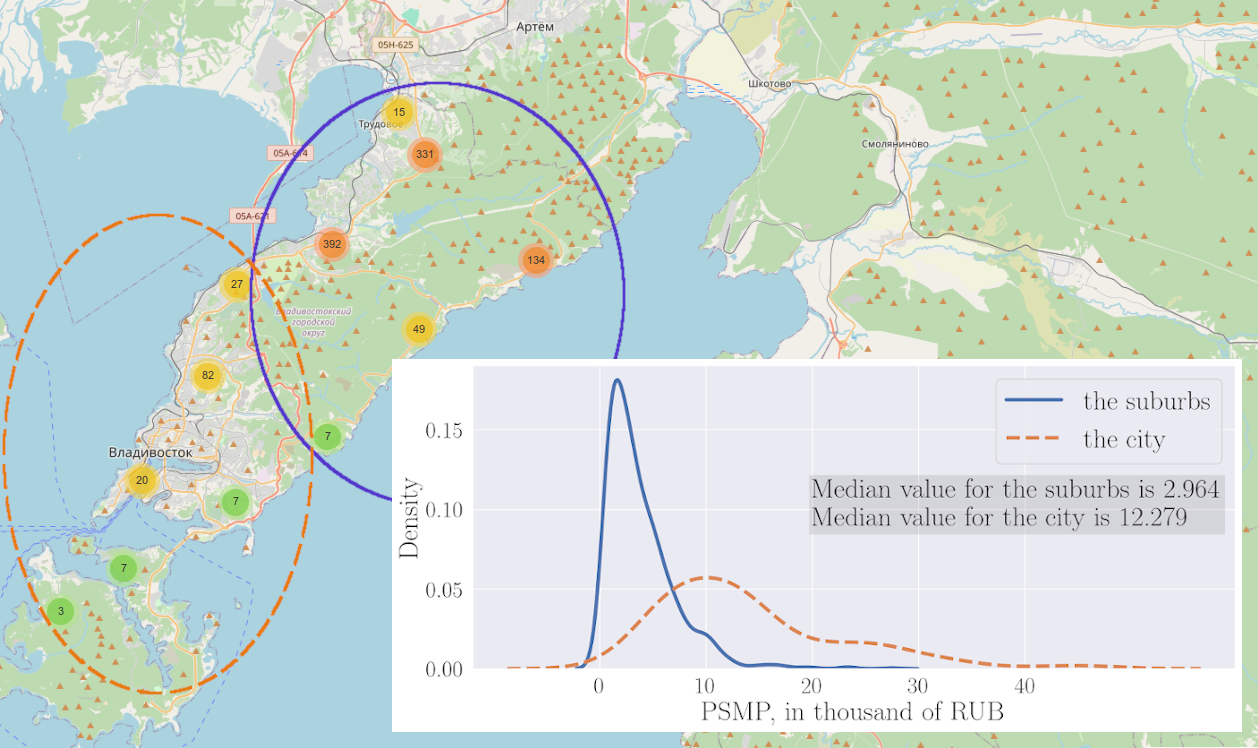}
    \caption{Comparison of median values of the per square meter price for the city and the suburbs (segment “Land parcels”). The values in filled circles show the number of objects for a particular area. The inset shows the estimations of the probability density functions (PDF) for PSMP. The triangular symbols represent hills and are added to the map automatically as part of the standard OpenStreetMap layout.}
    \label{fig:fig3}
\end{figure}

\begin{figure}[t]
    \includegraphics[width=\linewidth]{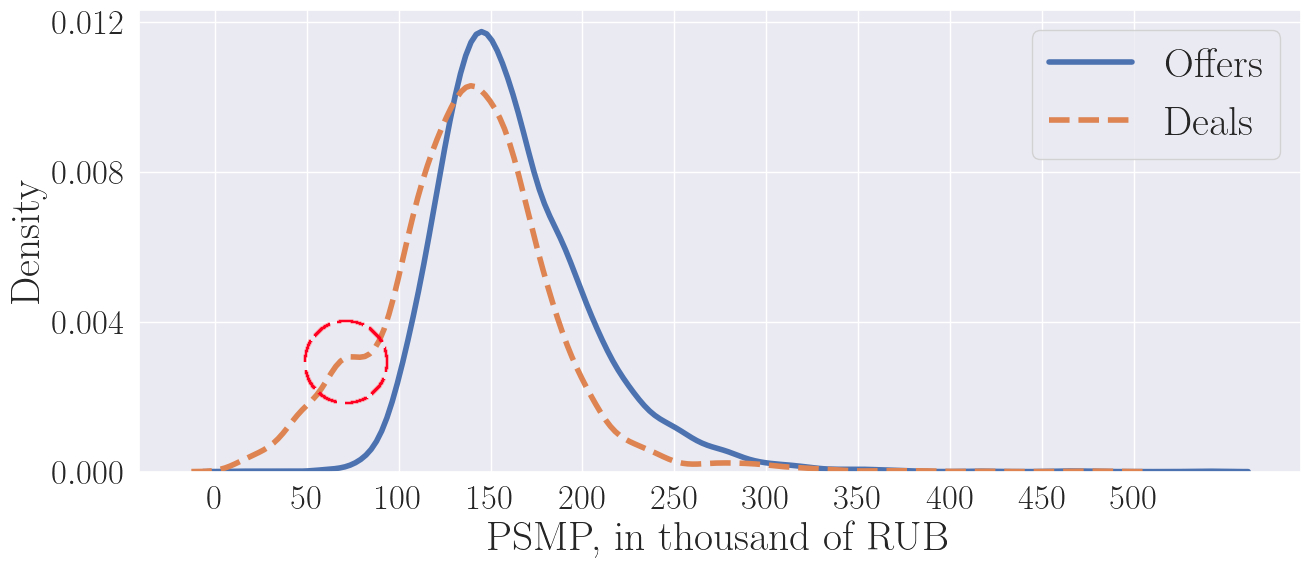}
    \caption{Comparison of probability density function for the offers and the deals (segment “Flats”, the whole dataset). The suspicious range of prices is shown by a dotted red circle. }
    \label{fig:fig4}
\end{figure}

Figure~\ref{fig:fig2} shows that points located close to each other can have significantly different the per square meter prices. This may be caused by individual characteristics of the object – the land parcel can be in a zone with a high risk of flooding or the land parcel can have a significant slope.
However, more likely, such difference in prices between neighboring land parcels means that one or both prices are anomalous and should be considered as outliers. Prices of objects like land parcels tend to change rather continuously with distance between them without sharp variations. Because of this reason, we can expect similar per square meter price among land parcels in some local area that changes with distance to other objects included in price factors (center of a town, water objects, social objects, etc.). 

For exclusion of outliers, we can start with clusterization of points on the base of their spatial relationships using, for example, k-means algorithm ~\citep{MacQuenn1967}. This method is simple for understanding and realization. It recognizes accumulation of points and creates clusters in such a way that the distance to the center of the cluster for every point is minimal, so neighboring points tend to fall into one cluster. We used SciKit-learn realization of this algorithm \citep{scikit-learn}. Next we can calculate the median or average value for each cluster and find the outliers using such methods as box-plot, z-score, and so on. Also, such clustering helps to take into account the inhomogeneity of an administrative territory. For example, the territory can include the city and the suburbs, where the per square meter prices differ significantly (Figure~\ref{fig:fig3}). On the contrary, if we consider the whole territory without clusterization we can lose part of the data because firstly the per square meter price in the city is much higher than the per square meter price in the suburbs and secondly the number of data points in the city is much less than the corresponding number in the suburbs. In this case, the land parcels located in the city become outliers, and by excluding them we lose the important data and significantly distort our predictions.

For the segment ``Flats'', we can find the outliers using the probability density function of the dependent variable. For example, Figure~\ref{fig:fig4} shows that values between 50000 RUB and 100000 RUB for the deals are doubtful.
We can check this assumption by inspecting individual buildings and analyzing the per square meter prices in each one. As an example, in Table~\ref{tab:tab1} we show the result of such inspection for one building highlighting the suspicious value. Such deals can be fictitious and negatively effect the forecast, so they should be excluded. 

\begin{table}[t]
  \begin{center}
    \caption{List of offers and deals (segment ``Flats'') for one particular building. A deal with the suspicious price is highlighted in red.}
    \label{tab:tab1}
    \begin{tabular}{|c|c|c|c|c|c|} 
    \hline
      \rowcolor{lightgray}
      \textbf{Type} &  \textbf{Storey} & \textbf{Total price, RUB} & \textbf{Area, m$^2$} & \textbf{PSMP, RUB}\\
      \hline
      offer & 5 & 3 850 000 & 17.1 & 225 146\\
      offer & 2 & 4 500 000 & 23.1 & 194 805 \\
      offer & 8 & 6 150 000 & 34.5 & 178 261 \\
      offer & 9 & 4 350 000 & 17.0 & 255 882 \\
      offer & 3 & 4 200 000 & 17.0 & 247 059 \\
      offer & 4 & 3 800 000 & 16.7 & 227 545 \\
      offer & 5 & 3 800 000 & 17.0 & 223 529 \\
      deal & 5 & 4 300 000 & 16.7 & 257 485 \\
      deal & 2 & 4 100 000 & 23.0 & 178 261 \\
      deal & 4 & 4 050 000 & 16.9 & 239 645 \\
      \rowcolor{pink}
      deal & 4 & 900 000 & 16.8 & 53 571 \\
      deal & 4 & 4 900 000 & 23.3 & 210 300 \\
      deal & 4 & 3 800 000 & 16.7 & 227 545 \\
      \hline
    \end{tabular}
  \end{center}
\end{table}

\begin{figure}[t]
    \includegraphics[width=\linewidth]{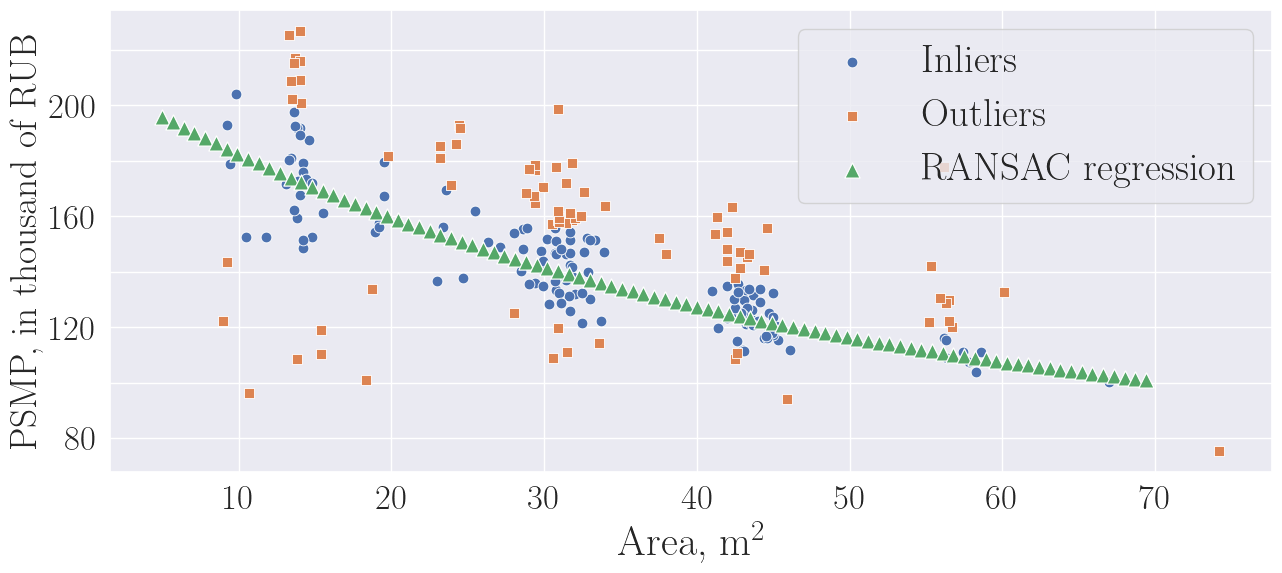}
    \caption{Results of the RANSAC regression for one of the subclusters of similar buildings formed by DBSCAN (segment “Flats”).}
    \label{fig:fig5}
\end{figure}

More serious methods such as machine learning also can be used for the identification of outliers. They are particularly helpful in the case of ``multi-dimensional'' objects like flats. First, we form territorial clusters in a similar manner as we did for the segment ``Land parcels''. Second, we create new subclusters in each territorial cluster on the base of the characteristics of the buildings such as wall material, number of storeys, year built, etc. This procedure allows to highlight typical buildings. We can identify such subclusters either manually on the basis of our expertise and knowledge of historic periods of city development, or we can use machine learning methods implementing clustering algorithms. Such algorithms take as input the buildings in the space of the above-mentioned characteristics and return the class markers for each building. In this paper, we use the Density-Based Spatial Clustering of Applications with Noise (DBSCAN) algorithm~\citep{Ester1996} with realization on Python (the library ``SciKit-learn''). This algorithm is based on the idea of closeness between points in the space of features. It means that DBSCAN groups neighboring points and marks single points as outliers. The minimal number of points in a neighborhood for a point to be considered as a core point, and the maximum distance between two points for one to be considered in the neighborhood of the other are adjustable parameters of the algorithm~\citep{Moreira2005}. The former controls the density of the resulting clusters. The latter is the most important one and should be chosen accurately depending on the data and the metric. The metric specifies how the distances in the space of features are calculated. After the formation of subclusters containing typical buildings, we start identification of outliers within each group. During the analysis, we noticed that there is a negative dependence between the per square meter price and the area of the flat. In addition, this dependence is not linear but rather exponential (Figure~\ref{fig:fig5}). Isolation of the trend for such data is a nontrivial task. If there are signs of linear dependence then we can try the RANdom Sample Consensus (RANSAC) regression~\citep{Kumar2020, Fischler1981} –- linear regression robust to outliers. This algorithm is an iterative procedure where every step assumes the formation of the subsample and fitting a linear model. Then this model is applied to the total sample and residuals are calculated. Points beyond the residual threshold are marked as outliers. Then the model retrains on another subsample and all steps are repeated. The new model is compared to the previous model with regard to the number of outliers. The model that contains fewer outliers is preferable. In our case, we can transform the exponential dependence to the linear by taking the natural logarithm of the data. So for each cluster found by DBSCAN, we apply the RANSAC regression to find the outliers and to isolate the trend. The dependent variable is the natural logarithm of the per square meter price and the explanatory variable is the area of a flat. This procedure allows to find the trend in the data with high fluctuations and noise. It hinges on a reasonable assumption that flats in buildings with similar characteristics located within some compact territory should have similar per square meter prices. The points that are highly different from the trend are considered to be outliers. These values can be explained by individual characteristics of a flat and should not be considered in the process of mass valuation. The results of the RANSAC regression for one of the subclusters are shown in Figure~\ref{fig:fig5}. Points marked as outliers are not used for forecasting. We used the median absolute deviation (MAD) as the threshold, which is the default option in the ``SciKit-learn'' library.

\section{Formation and selection of explanatory variables}
\label{select_variables}

\subsection*{Initial collection of features}
\label{initial_collection}
According to the methodological rules of mass valuation of real estate~\citep{Method2018}, we can use as explanatory variables such information as general characteristics of an object (object name, purpose, area, wall material, year built, landscape, kind of soil, and so on); information about location of an object (distance to roads, water objects, recreation areas, city center, and so on); information about engineering infrastructure (class of a linear object, distance to a power line, rated power of a linear object, and so on); other matter (price level of a commodity bundle, commodity circulation per person, distance to extractive industry territory, and so on).

\begin{figure}
    \includegraphics[width=\linewidth]{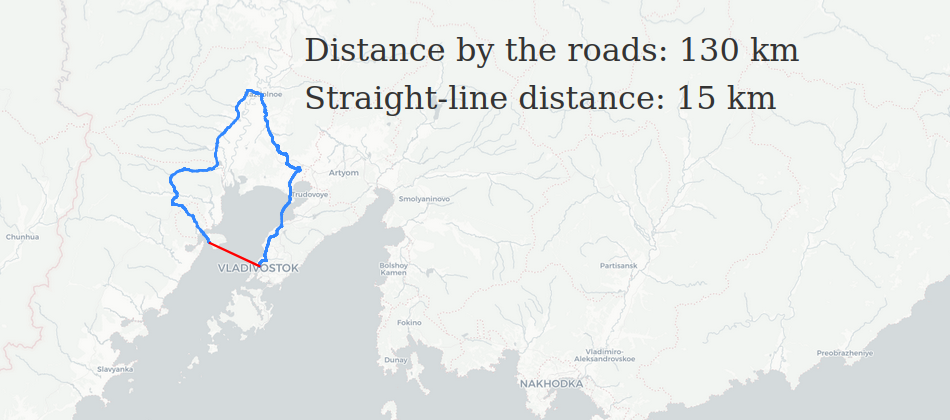}
    \caption{Comparison of the straight-line distance and distance by the roads between the center of Vladivostok and Beregovoye which is the part of Vladivostok within Frunzensky district.}
    \label{fig:fig6}
\end{figure}

General information about the object is contained in the Unified State Register of Real Estate and in accordance with the law is provided by the ministry to the certified organization that makes evaluation. Socioeconomic data can be found in the annual reviews of the territory development. Geographical features can be extracted using commercial mapping software like NextGIS\footnote{\url{https://nextgis.com}} and ArcGIS\footnote{\url{https://www.esri.com/software/arcgis}}, or nonprofit resources like OpenStreetMap~\citep{OSM} facilitated by the Python library “OSMnx”~\citep{Boeing2025}. It provides information on all kinds of socially significant facilities –- nursery schools, schools, hospitals, shops, malls, and so on. Then the straight-line distances to these objects or the number of such objects in some radius can be used as explanatory variables for a predictive model. Additionally, libraries “OSMnx” and “NetworkX”~\citep{NetworkX} allow to load road network graphs providing the opportunity to calculate distances by roads which improves the accuracy of the variables. It is important for territories with irregular coastlines (Figure~\ref{fig:fig6}), mountain ranges, or isolated territories like islands. In the last case, we can use adjusting factors reflecting the accessibility of the island.

\subsection*{Construction of new features}
\label{construction_features}
We can aggregate some variables into one (as explained below in the discussion of correlation matrices) or use conceptions from related fields, for example, graph theory. Thus, we created the factor “Development of the road network” that helped to estimate the location of an object in relation to transport accessibility. If the object has many crossroads and roads nearby, then this object is more attractive and therefore more expensive. This idea is especially important for big cities. For derivation of the factor “Development of the road network” we used the road network graphs and the conception of centrality which allows to find the most important nodes of a graph~\citep{Newman2010}. To calculate the centrality, we consider crossroads and endpoints of roads as nodes of a graph and roads between these points as graph edges. In these terms the importance of a node means how many shortest roads between two points go through this node. In order to calculate the length of a route, we naturally associate the road distance between two points with the weight of the corresponding graph edge. The shortest route between two points is the one for which the sum of weights of constituting edges is minimal. The Python library “NetworkX” allows to calculate the centrality in different ways. In this paper, we use the harmonic centrality~\citep{Marchiori2000} because of its simplicity and straightforward interpretation. The harmonic centrality of a node is the sum of the inverse lengths of the shortest routes from each node to this node. It has a perfect intuitive interpretation –- smaller distances give a higher contribution to centrality. Mathematically, the harmonic centrality can be expressed as

\begin{equation}
    C(u) = \sum_{v \neq u} \frac{1}{d(v,u)},
    \label{eq:harmonic_centrality}
\end{equation}
where $u$ is the node for which the centrality is calculated, $v$ is some other node, $d(v,u)$ is the length of the shortest route between $u$ and $v$.

\begin{figure*}[!htb]
    \centering
\begin{tabular}{cc}
	        \includegraphics[width=0.48\textwidth]{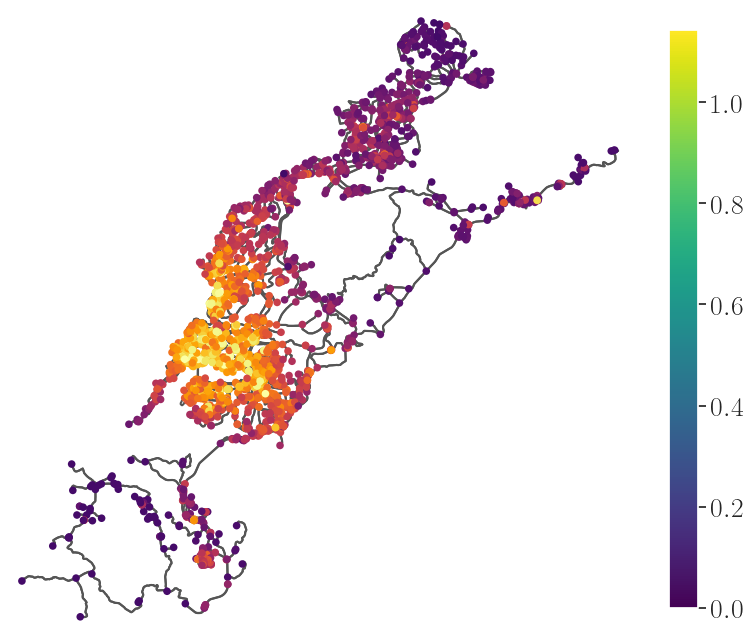}
	    &
	        \includegraphics[width=0.48\textwidth]{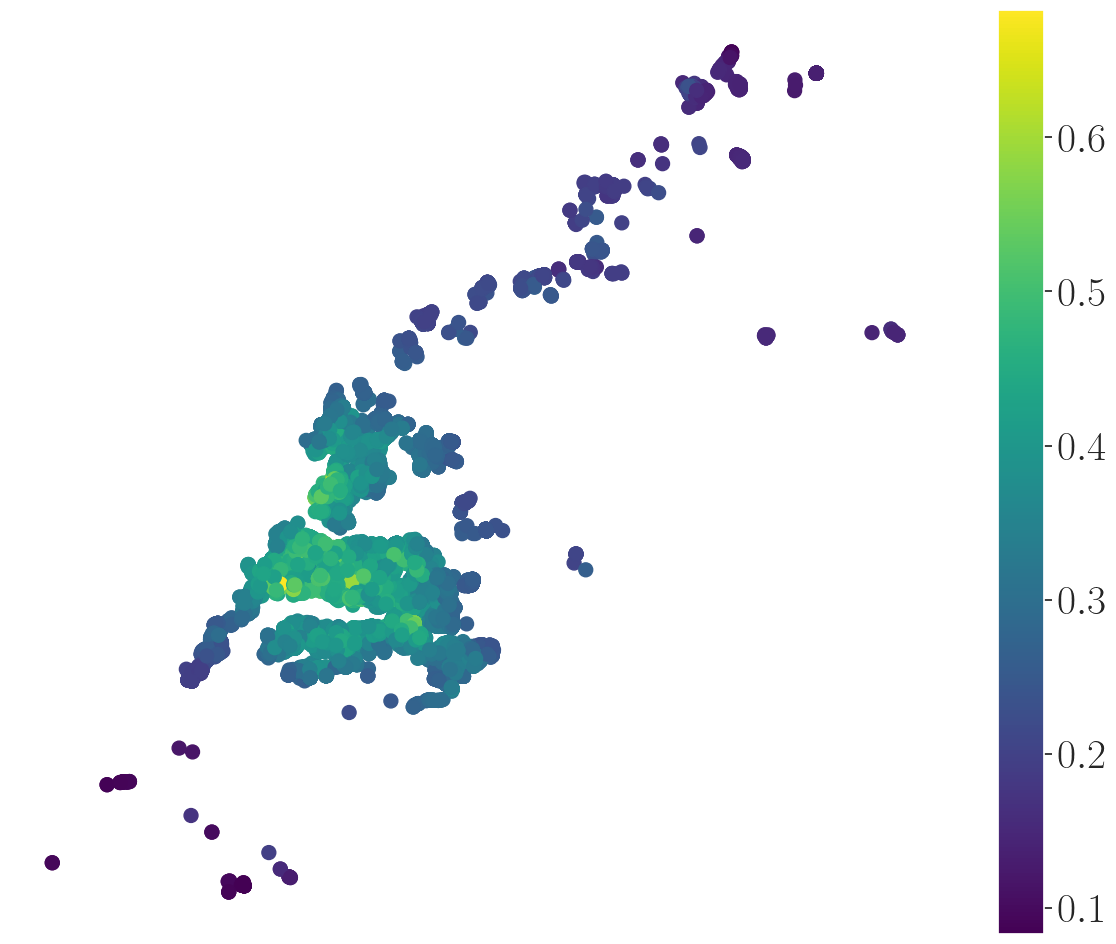}
	     \\
	        (a)
	    &
	        (b)
\end{tabular}
	\caption{\textbf{(a)}: Centrality of graph nodes corresponding to road network of Vladivostok. \textbf{(b)}: Factor “Development of the road network” for the training data set (segment “Flats”).}
	\label{fig:fig7}
\end{figure*}

The roads can have bidirectional and unidirectional traffic. Since the graph of the road network is an oriented graph, every edge of the graph has a direction, so inversion of end points of the route potentially can change its length. This fact should be taken into account in calculation of the centrality. Figure~\ref{fig:fig7} (a) shows the harmonic centrality for the road network of Vladivostok.

The mapping between the centrality of nodes and the centrality of objects for evaluation can be done in several ways ordered from the simplest to more sophisticated:

\begin{enumerate}
  \item  
  Find the closest to the object node of the road network; consider its centrality as an estimation of the centrality of the object.
  \item 
  Create the buffer area of some radius around the object; find the nodes of the road graph that lie inside this area; calculate the mean value of the centrality among these nodes; consider this result as an estimation of the centrality of the object. 
  \item Calculate the centrality for all nodes of the road graph; interpolate over nodes to get a smooth continuous surface of centrality values. 
\end{enumerate}

We implement the third approach because it allows to estimate the centrality of objects more precisely compared to other mentioned possibilities. We perform the interpolation using the method of basis functions in which the target variable is expressed as a linear combination of some functions $B(h_{0i})$ that depend on the distance~\citep{Demyanov2010,Buhmann2000,Regis2006,Zhang2024}, 

\begin{equation}
    Z(x_0) = \sum_{i=1}^{N}c_iB(h_{0i}),
    \label{eq:basis_func}
\end{equation}
where $Z(x_0)$ is the estimation of centrality at the point $x_0$, $h_{0i}$ is the distance from point $x_0$ to point $x_i$, $c_i$ are the weight coefficients defining the algebraic sign of an element and its contribution.
As the basis functions we use the linear functions of the form

\begin{equation}
   B(h) = -h.
   \label{eq:basis_func_linear_spline}
\end{equation}

The weight coefficients $c_i$ are derived from the condition that at each graph node \((x_i,y_i)\) the result of interpolation must coincide with this node centrality value \(V(x_i,y_i)\). Thus, we need to solve the system of $N$ linear equations with $N$ unknown variables, 

   \begin{equation}
    \begin{cases}
      \sum_{i=1}^{N}c_iB(h(x_i,y_i,x_1,y_1)) = V(x_1,y_1), \\
      ... \\
      \sum_{i=1}^{N}c_iB(h(x_i,y_i,x_N,y_N)) = V(x_N,y_N).
    \end{cases}\
\end{equation}
    \label{eq:basis_func_linear_spline_sys}

We perform the interpolation using the RBFInterpolator from Python library ``SciPy''~\citep{Virtanen2020,ScipyDoc}. The results for Vladivostok are shown in Figure~\ref{fig:fig7} (b). The factor ``Development of the road network'' essentially represents the estimated centrality of the objects.

\begin{figure*}[t]
    \centering
\begin{tabular}{cc}
	        \includegraphics[width=0.48\textwidth]{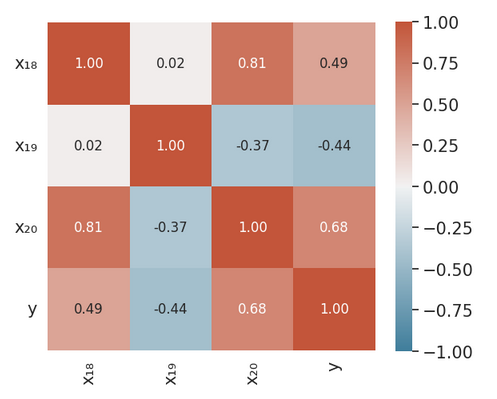}
	    &
	        \includegraphics[width=0.48\textwidth]{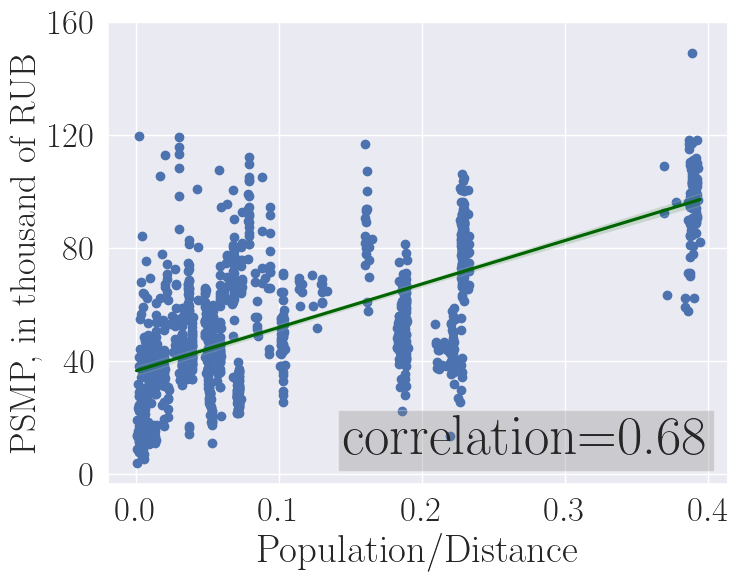}
	     \\
	        (a)
	    &
	        (b)
\end{tabular}
	\caption{\textbf{(a)}: Correlation matrix for features ``$x_{18}$: Population of a settlement'', ``$x_{19}$: Distance to Vladivostok (by roads)'', ``$x_{20}$: Population of a settlement divided by Distance to Vladivostok (by roads)'', ``$y$: Per square meter price'' (segment ``Flats''). \textbf{(b)}: The relation between the value of the per square meter price and the aggregated factor $x_{20}$ (segment ``Flats''). The best fit line is shown in green.}
	\label{fig:fig8}
\end{figure*}

In the case of an inhomogeneous area (e.g. if the territory contains a few towns or villages, like parts of Primorsky Krai near Vladivostok) then the model can be improved by aggregation of the existing variables into new ones. For example, from Figure~\ref{fig:fig8} (a) one can see that independent variables ``$x_{18}$: Population of a settlement'' and ``$x_{19}$: Distance to Vladivostok (by roads)'' have quite significant correlation of about 0.5 with the dependent variable ``$y$: Per square meter price''.
On the one hand, increasing of the distance between a settlement and Vladivostok administration negatively affects the per square meter price. On the other hand, increasing of the population of a settlement positively affects the per square meter price. These considerations led us to create a new factor that aggregates these two variables -- “\(x_{20}\): Population of a settlement divided by the distance to Vladivostok by roads”. Econometrically, this factor should have a positive correlation with the per square meter price. Figure~\ref{fig:fig8} (b) shows the relation between this factor and the value of per square meter price of a flat. The resulting correlation between this factor and the dependent variable is significantly higher comparing to \(x_{18}\) and \(x_{19}\).

We described a couple examples of construction of new features relevant for real market prices prediction. Further information and ideas can be found in e.g. ~\cite{Galli2020}.

\subsection*{Selection of features}
\label{select_features}

The next important step on the way of creating the model is estimation of correlations of all factors with the dependent variable (the per square meter price) and selection of the most important of them. At this step, we created the correlation matrix for quantitative factors and we analyzed probability density functions of the per square meter price for categorical factors.

\begin{figure}[t]
    \includegraphics[width=\linewidth]{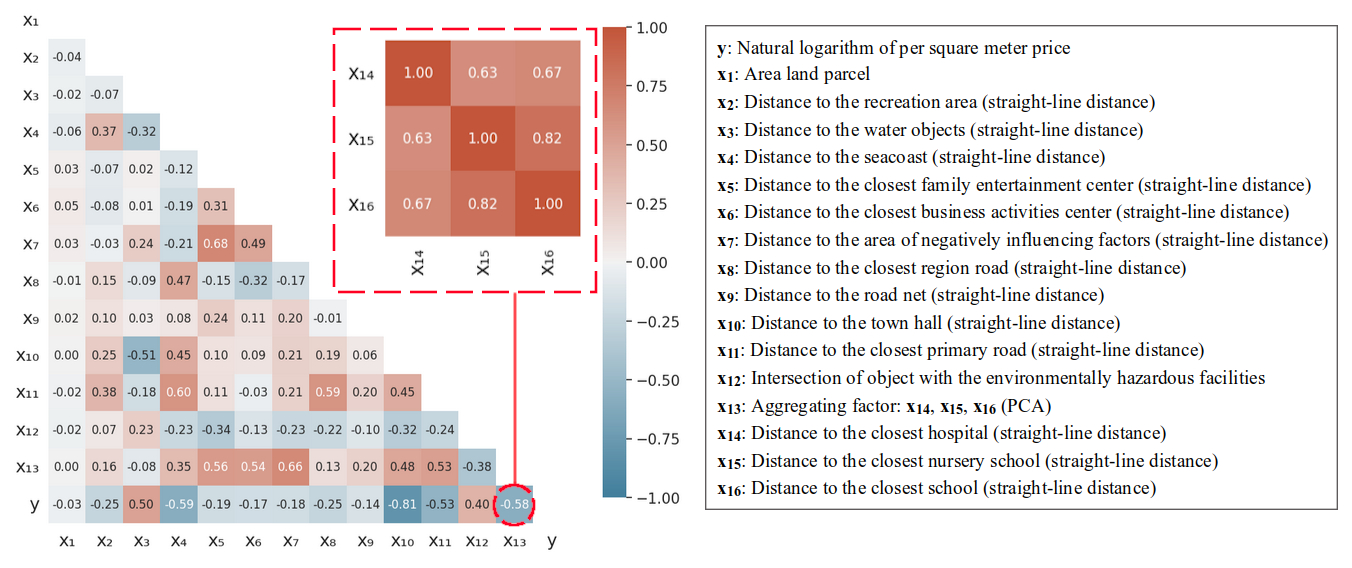}
    \caption{Correlation matrix after exclusion of outliers and multicollinearity (segment “Land parcels”).}
    \label{fig:fig9}
\end{figure}

\begin{figure}
    \includegraphics[width=\linewidth]{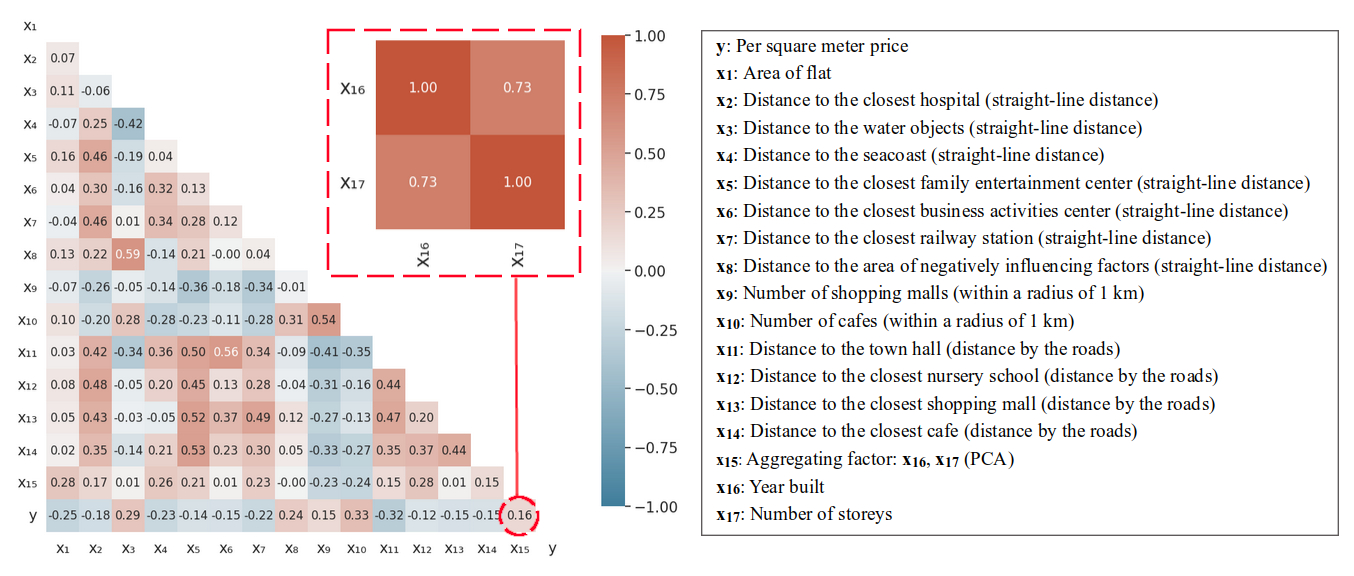}
    \caption{Correlation matrix after exclusion of outliers and multicollinearity (segment “Flats”).}
    \label{fig:fig10}
\end{figure}

When we study the correlation matrices (Figure~\ref{fig:fig9}, Figure~\ref{fig:fig10}) we are interested not only in the relationships of the explanatory variables with the dependent variable but also in the correlations between explanatory variables themselves. Too high correlation \citep[$r \gtrsim 0.7$,][]{Kozak2009,Gujarati2009} between the explanatory variables indicates the presence of multicollinearity in the data. Since we base our modeling on the linear regression, multicollinearity should be avoided~\citep{Hadi2012}. We can get rid of the multicollinearity in the data in several ways: keep only the factors that have higher correlation with the dependent variable; drop the factors with the wrong sign of correlation with the dependent variable indicating that such factors have incorrect economic substance; integrate several factors into one employing the principal component analysis \citep[PCA, ][]{Ajvazyan2010}. PCA is useful when several factors have similar economic substance. For example, the distance to a school, a nursery school, and supermarkets can be aggregated into one factor which is the distance to socially important facilities (alternatively, instead of the distance we can use the number of objects). Such aggregation is justified by the fact that these features have significant correlation with each other (Figure~\ref{fig:fig9}), but if we drop one of them we lose some information. Instead, we incorporate them into the principal component that includes the dispersion of all these factors allowing us to reduce the number of features while keeping as much information as possible. It is worth noting that the majority of the considered features reflect spatial characteristics of the data. Inclusion of such features is crucially important in building of an effective model of spatial data~\citep{Jahanshiri2011} like land parcels and flats.

\begin{figure}
    \includegraphics[width=\linewidth]{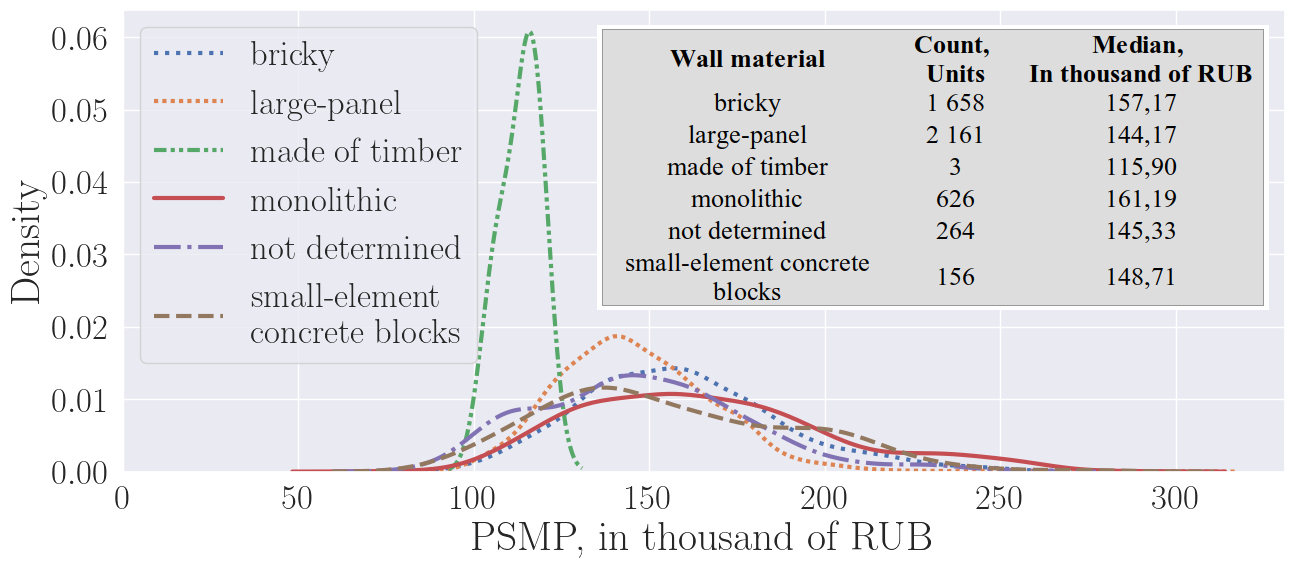}
    \caption{Comparison of probability density functions for the factor “Wall material” (segment “Flats”).}
    \label{fig:fig11}
\end{figure}

Previously we noted that before including categorical factors in the model we plot probability density functions of the per square meter price. A similar method of analysis would be drawing of a box-plot diagram~\citep{Tukey2000}. Traditionally, categorical factors are included in a model as binary features and one should pay attention not only to the average values of the per square meter price for each group but also to the representativity of the group in the data. For example,  Figure~\ref{fig:fig11} shows the probability density function for the factor “Wall material”. There are only 3 flats in timber buildings, and while their per square meter price is far less than for the other materials, we can not include this binary feature in the model due to its non-representativity. 

After completing all preparative steps –- removing multicollinearity, dropping the factors without the economic substance, and dropping the non-significant features, we can move to selection of the price-forming factors that should be included in the model.
There are different approaches~\citep{Masis2021} to this problem which generally can be grouped into three classes~\citep{fea_sel_guide}: filter methods, wrapper methods, and embedded methods. The first are independent of any machine learning algorithms. Instead, features are selected on the basis of their scores in various statistical tests on their correlation with the outcome variable. The library ``SciKit-learn'' offers SelectKBest procedure that takes two arguments: the number $k$ of the most significant features and the type of the statistical test. The examples are ANOVA F-value test or mutual information test\footnote{\url{https://scikit-learn.org/stable/modules/feature\_selection.html}}. The second class, wrapper methods, selects a subset of features and trains a model on them. Features are added or removed based on the performance of the model and/or their importance.
The selection of features is essentially reduced to a search problem. This approach is usually very expensive in terms of computational resources. The examples are exhaustive feature selection (considers all possible subsets of features), forward selection (starts with zero features consequently adding features which best improve the model until addition of new factors does not improve the performance of the model), and backward elimination (opposite to forward selection, starts with the whole set of features and at each iteration removes a feature the least contributing to the model performance). The typical performance metrics are \(R^2\), mean absolute error, mean squared error, median absolute error (deviation), etc. The third class, the embedded methods, consists of models with penalty regularization. These models contain the additional term penalizing them for inclusion of too many features. The classical example is the Ridge regression (also known as L2 regularization) introduced by~\cite{Hoerl1970} with the error function
\begin{equation}
    J = \sum_i(y_i-\hat{y}_i)^2 + \lambda\omega^2,
    \label{eq_ridge}
\end{equation}
where $y_i$ is the actual value, $\hat{y}_i$ is the predicted value, $\omega$ is the vector of coefficients, and $\lambda$ is the hyperparameter which can be adjusted manually in order to control the penalty on the number of coefficients and their magnitudes.
Another instance of this approach is the LASSO (Least Absolute Shrinkage and Selection Operator) regression~\citep{Tibshirani1996} also known as L1 regularization with the error function
\begin{equation}
    J = \sum_i(y_i-\hat{y}_i)^2 + \lambda||\omega||.
    \label{eq_lasso}
\end{equation}
The crucial difference between these two regularizations is that the former decreases some coefficients to the values around zero while the latter shrinks them to exactly zero values facilitating the exclusion of corresponding features from the model. We discuss the consequences in Section~\ref{modeling}. The combination of these two approaches is known as the Elastic net~\citep{Zou2005} and contains both L1 and L2 norms in the error function. All these models can be reduced to ordinary least squares by setting hyperparameter $\lambda$ to zero. While embedded methods are a powerful technique, in terms of interpretability it is better first to try filter and wrapper methods because they provide more flexibility for modeling: the stages of feature selection and model construction are separated allowing an appraiser to perform fine adjustment of the model. This reasoning is valid in the case of a reasonable number of features. When one has to deal with tens or even hundreds of features, attractiveness of embedded methods sufficiently increases (see Section~\ref{modeling}).

\begin{table}[t!]
  \begin{center}
    \caption{Comparison of results of linear regression using Recursive Feature Elimination (segment “Land parcels”) depending on the number of factors included in the model. The optimal number of features is highlighted in green.}
    \label{tab:tab2}
    \begin{tabular}{|C|c|c|c|c|} 
    \hline
      \rowcolor{lightgray}
      \textbf{Number of factors} & \textbf{$R_\mathrm{adj}^2$} &  \textbf{Durbin-Watson} & \makecell{\textbf{$p$-value}\\ \textbf{(Jarque-Bera)}} & \makecell{\textbf{$p$-value}\\ \textbf{($F$-statistic)}} \\
      \hline
      1 & 0.634 & 1.238 & 0.000005 & $\ll$ 0.001 \\
      2 & 0.712 & 1.371 & 0.073 & $\ll$ 0.001 \\
      3 & 0.749 & 1.564 & 0.175 & $\ll$ 0.001 \\
      \rowcolor{SeaGreen}
      4 & 0.753 & 1.564 & 0.074 & $\ll$ 0.001 \\
      5 & 0.763 & 1.610 & 0.003 & $\ll$ 0.001 \\
      6 & 0.773 & 1.567 & 0.004 & $\ll$ 0.001 \\
      \hline
    \end{tabular}
  \end{center}
\end{table}

For the segment “Land parcels”, where the number of features is not very large and their correlations with the dependent variable have sufficiently high values (Figure~\ref{fig:fig9}), we applied the Recursive Feature Elimination (RFE) algorithm~\citep{Guyon2002} which is another example of wrapper methods. This algorithm starts with the full set of features, and consequently removes one feature at each iteration. Conceptually it is very similar to Backward elimination, but the mechanism is different. There are two popular realizations of RFE and they differ substantially from each other. The SciKit-learn version\footnote{\url{https://scikit-learn.org/stable/modules/generated/sklearn.feature_selection.RFE.html}} at each iteration builds the model and ranks features by their importance on the base of the corresponding $p$-values (for linear models) and excludes the least significant one. It does not evaluate the performance of the model. The Feature-engine version\footnote{\url{https://feature-engine.trainindata.com/en/latest/user_guide/selection/RecursiveFeatureElimination.html}} is some sort of a hybrid between SciKit-learn realization and Backward elimination. At each iteration it evaluates the performance of the model and ranks features by their importance. Next, it removes the least significant feature and evaluates the model again. If it scores less than the previous model then the excluded feature is returned back to the model and this step is repeated for the second worst feature, and so on. If removing the feature improves the performance of the model then the remaining features are ranked again and the process is repeated. In our study we use the SciKit-learn version taking into account the value of $R^2_\mathrm{adj}$~\citep{Gujarati2009} and checking fulfillment of the basic conditions for the ordinary least squares method~\citep{Ajvazyan2010,Kutner2005}. Such conditions are: the residuals should not have significant autocorrelation \citep[i.e. the Durbin–Watson statistic should be in the range from 1.5 to 2.5, where the value 2.0 means that the residuals are completely uncorrelated, ][]{Field2012}; the residuals should be normally distributed (i.e. they should pass the Jarque-Bera test with $p$-value (JB) $>$ 0.05~\citep{Gujarati2009} which corresponds to non-rejection of the null hypothesis of normality); the equation of regression should be significant with $p$-value ($F$-statistic) $<$ 0.01~\citep{Gujarati2009}, corresponding to rejection of the null hypothesis that all coefficients of the model except the intercept are zero.

\begin{table}[t]
  \begin{center}
    \caption{Same as Table~\ref{tab:tab2} but for the segment “Flats”. Too low $p$-values in the Jarque-Bera test (highlighted in red) indicate violation of the requirement of the residuals normality. RFE is not applieble regardless of the number of factors.}
    \label{tab:tab2_1}
    \begin{tabular}{|C|c|c|>{\columncolor{pink}}c|c|} 
    \hline
      \rowcolor{lightgray}
      \textbf{Number of factors} & \textbf{$R_\mathrm{adj}^2$} &  \textbf{Durbin-Watson} & \makecell{\textbf{$p$-value}\\ \textbf{(Jarque-Bera)}} & \makecell{\textbf{$p$-value}\\ \textbf{($F$-statistic)}} \\
      \hline
      1 & 0.200 & 2.036 & $\ll$ 0.001 & $\ll$ 0.001 \\
      2 & 0.266 & 2.040 & $\ll$ 0.001 & $\ll$ 0.001 \\
      3 & 0.303 & 2.047 & $\ll$ 0.001 & $\ll$ 0.001 \\
      4 & 0.397 & 2.059 & $\ll$ 0.001 & $\ll$ 0.001 \\
      5 & 0.467 & 2.081 & $\ll$ 0.001 & $\ll$ 0.001 \\
      6 & 0.499 & 2.063 & $\ll$ 0.001 & $\ll$ 0.001 \\
      \hline
    \end{tabular}
  \end{center}
\end{table}

Table~\ref{tab:tab2} shows that in the case of ``Land Parcels'' four factors are the optimal choice taking into account the aforementioned requirements. Three factors also satisfy the statistical tests and have a similar $R^2_\mathrm{adj}$, but all other things being equal retaining more factors improves the descriptive power of the model.

For the segment ``Flats'' of RFE performs poorly (Table~\ref{tab:tab2_1}). Small values of $R^2_\mathrm{adj}$ can be explained by significantly lower correlation of features with the dependent variable (Figure~\ref{fig:fig10}) compared to the segment ``Land parcels'' (Figure~\ref{fig:fig9}). But the biggest problem is the near-zero $p$-values in the Jarque-Bera test, which means that the residuals of the models are not distributed normally. The SelectKBest algorithm also failed to find a consistent subset of features. The inability of different algorithms to build a linear model along with small $R^2_\mathrm{adj}$ and correlations strongly suggests that other models should be considered. In this work we used the Random Forest algorithm to generate rules from existing features with further substitution of created rules as new features to L1 regularization, which provides better results without violating the interpretability constraint. It is presented in details in Section~\ref{modeling}.

Recursive Feature Elimination is the main technique used for feature selection in our study. It leads to sound linear model for the segment ``Land Parcels'' which is described in the next section. For segment ``Flats'' no discussed methods allows building the ordinary least squared regression, so its advanced variations are considered. Comparative analysis of various feature selection methods for the particular dataset is beyond the scope of this paper. The examples of research addressing this topic in the context of regression models describing economic and environmental data are \cite{Banga2021,Jomthanachai2022,Mohapatra2022}.

\section{Modeling}
\label{modeling}

In this paper, we show the results of prediction of the cadastral value for objects located in Primorsky Krai, Russia. The region has a capital city – Vladivostok (with a population of about 600 thousand people), towns (with a population of not more than 150 thousand people), and villages. All these territories have different property markets, price levels, and quality of life, which depend on economic and territorial characteristics. The methodological rules~\citep{Method2018} recommend splitting the territory on the base of administrative boundaries of municipalities, taking into account the population. This recommendation can be applied to objects that are located on homogeneous territory, e.g. flats located in one city. However, it works poorly for objects that are spread over a large area, e.g. land parcels. They occupy a large heterogeneous area of the city and the suburbs, and often form allotment societies (also called gardeners' non-commercial partnerships). Gardeners' non-commercial partnerships are a collective form of gardening. Members perform their activities on an individual basis. Activities of such partnerships are regulated by the federal law. The partnerships can have their regulations, convoke meetings, and appoint a chairman. Similar organizations exist in many countries, for example, Germany, Sweden, Poland, and the USA. In every country, there are its own procedures for registration and obligation to pay taxes. In Russia, such partnerships must pay taxes, in particular the land tax~\citep{Seregina2019,Bykowa2024} –- the tax paid by individuals or legal entities that have the property right of land plot or the right of land plot perpetual usage or the right of lifetime inheritable possession. Tax amount is defined from the cadastral value and taxation rate. Such allotment societies can include different administrative regions. In such cases, it is better to use a grouping of objects employing the clusterization algorithms instead of using formal administrative borders. Most of the algorithms are based on spatial relations between objects. Because of this reason, for the segment “Land parcels” we used a k-means clustering algorithm, that takes into account spatial positions of objects.

All further calculations were done for the capital region: Vladivostok and its suburbs.

\subsection*{segment “Land parcels”} 

\begin{figure*}[t]
    \centering
\begin{tabular}{cc}
	        \includegraphics[width=0.48\textwidth]{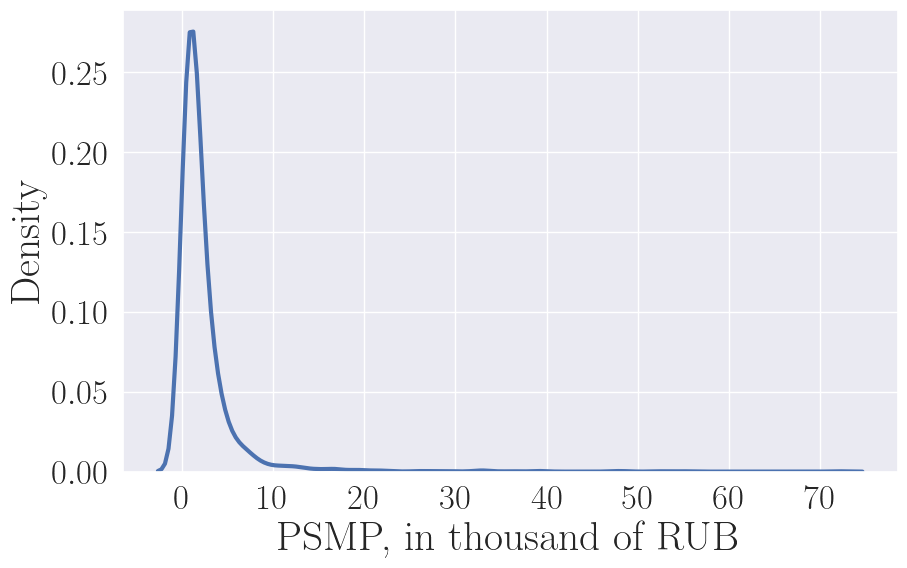}
	    &
	        \includegraphics[width=0.475\textwidth]{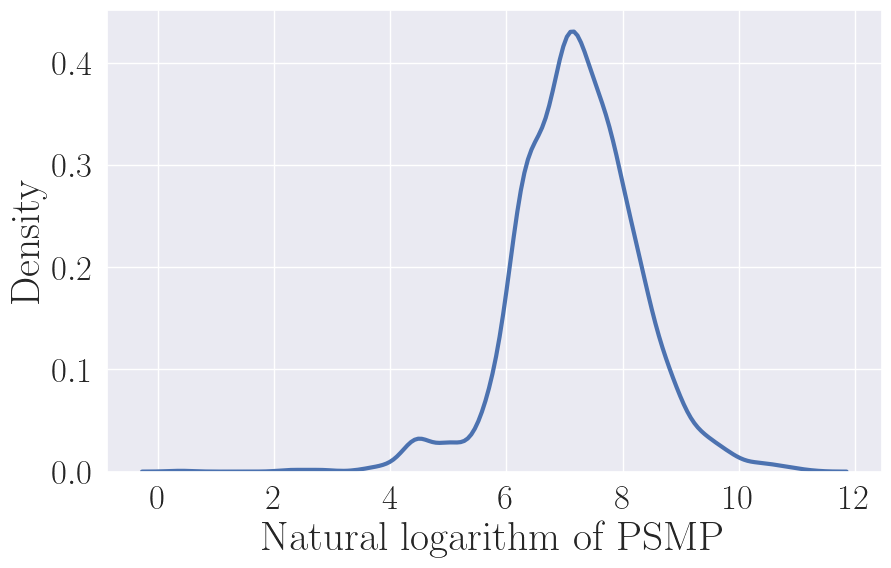}
	     \\
	        (a)
	    &
	        (b)
\end{tabular}
	\caption{Comparison of  the probability density function of the dependent variable \textbf{(a)} with its logarithm \textbf{(b)} (segment “Land parcels”).}
	\label{fig:fig12}
\end{figure*}

From the plot of the probability density function of the dependent variable (per square meter price) shown in Figure~\ref{fig:fig12} (a) one can see that it is not normal.
Although there is no requirement on the normality of the dependent variable, analysts tend to prefer working with the data distributed approximately normally.
Because of this reason, we used a natural logarithm of the per square meter price (Figure~\ref{fig:fig12}, b). Also, such transformation of data guarantees that prediction of modeling will have only positive values.

\begin{table}[!t]
    \caption{Results of training of the linear regression (segment “Land parcels”).}
    \label{tab:tab3}
    \begin{tabular}{|>{\centering\arraybackslash}m{3cm}>{\centering\arraybackslash}m{1.95cm}>{\centering\arraybackslash}m{1.95cm}>{\centering\arraybackslash}m{1.95cm}>{\centering\arraybackslash}m{1.95cm}|}
     \hline
     \multicolumn{5}{|c|}{\textbf{OLS Regression Results}} \\
     \hline
     \multicolumn{1}{|c}{Dep. variable:} & \multicolumn{4}{c|}{$y$} \\
     \multicolumn{5}{|c|}{} \\
     \multicolumn{1}{|c}{$R^2$:} & \multicolumn{4}{c|}{0.761} \\
     \multicolumn{1}{|c}{$R_\mathrm{adj}^2$:} & \multicolumn{4}{c|}{0.760} \\
     \multicolumn{1}{|c}{$F$-statistic:} & \multicolumn{4}{c|}{740.8} \\
     \multicolumn{1}{|c}{$p$-value ($F$-statistic):} & \multicolumn{4}{c|}{$\ll$ 0.001} \\
     \hline
      & \centering\textit{coef} & \centering\textit{std err} & \centering\textit{t} & $P>|t|$ \\
     \hline
     \centering{$const$} & \centering{7.3210} & \centering{0.012} & \centering{627.655} & $\ll 0.001$ \\
     \centering{$x_3$} & \centering{0.1070} & \centering{0.014} & \centering{7.585} & $\ll 0.001$ \\
     \centering{$x_4$} & \centering{-0.1642} & \centering{0.013} & \centering{-12.254} & $\ll 0.001$ \\
     \centering{$x_{10}$} & \centering{-0.3719} & \centering{0.016} & \centering{-22.918} & $\ll 0.001$ \\
     \centering{$x_{13}$} & \centering{-0.1153} & \centering{0.009} & \centering{-12.876} & $\ll 0.001$ \\
     \multicolumn{5}{|l|}{$y$ is Natural logarithm of the per square meter price} \\ 
     \multicolumn{5}{|l|}{$x_3$ is Distance to the water objects (straight-line distance)} \\ 
     \multicolumn{5}{|l|}{$x_4$ is Distance to the seacoast (straight-line distance)} \\ 
     \multicolumn{5}{|l|}{$x_{10}$ is Distance to the town hall (straight-line distance)} \\ 
     \multicolumn{5}{|l|}{$x_{13}$ is Aggregating factor: $x_{14}$, $x_{15}$, $x_{16}$ (PCA)} \\ 
     \multicolumn{5}{|l|}{$x_{14}$ is Distance to the closest hospital (straight-line distance)} \\ 
     \multicolumn{5}{|l|}{$x_{15}$ is Distance to the closest nursery school (straight-line distance)} \\ 
     \multicolumn{5}{|l|}{$x_{16}$ is Distance to the closest school (straight-line distance)} \\
     \hline
     \multicolumn{1}{|c}{Durbin-Watson:} & \multicolumn{4}{c|}{1.572} \\
     \multicolumn{1}{|c}{Jarque-Bera (JB):} & \multicolumn{4}{c|}{3.713} \\
     \multicolumn{1}{|c}{$p$-value (JB):} & \multicolumn{4}{c|}{0.156} \\
     \multicolumn{1}{|c}{Cond. No.:} & \multicolumn{4}{c|}{2.87} \\
     \hline
    \end{tabular}
\end{table}

We start the modeling with the simplest interpretable model –- the linear regression, using the subset of four selected features as described in Section~\ref{select_features}. The data was split into train/test parts with proportion 70\%/30\%. The ``statsmodels''~\citep{seabold2010statsmodels} library was used to train the model. The results are summarized in Table~\ref{tab:tab3}.

\begin{table}[!t]
  \begin{center}
    \caption{VIF for factors included in the linear regression (segment “Land parcels”).}
    \label{tab:tab4}
    \begin{tabular}{|>{\centering\arraybackslash}m{0.85\linewidth}|>{\centering\arraybackslash}m{0.08\linewidth}|} 
    \hline
      \rowcolor{lightgray}
      \textbf{Factors} & \textbf{VIF} \\
      \hline
      Distance to the water objects (straight-line distance) & 1.463 \\
      Distance to the seacoast (straight-line distance) & 1.320 \\
      Distance to the town hall (straight-line distance) & 1.936 \\
      Aggregate: Distance to the hospitals (straight-line distance) \text{ \& } Distance to the closest nursery school (straight-line distance) \text{ \& } Distance to the closest school (straight-line distance) & 1.424 \\
      \hline
    \end{tabular}
  \end{center}
\end{table}

\begin{figure}[!t]
    \includegraphics[width=\linewidth]{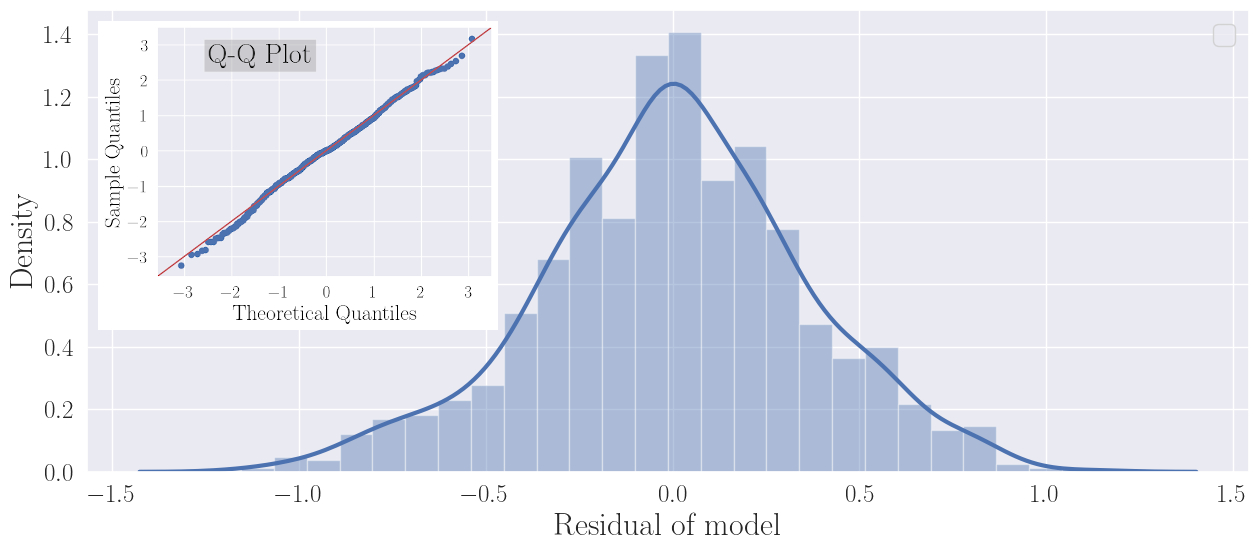}
    \caption{Distribution of the linear model residuals (segment “Land parcels”). The inset shows Quantile-Quantile plot to compare quantiles of model residuals to quantiles of the normal distribution. All points arrange along the diagonal line showing that the distribution of residuals approaches the normal distribution.}
    \label{fig:fig13}
\end{figure}

As shown in Table~\ref{tab:tab2} from the previous section, the model satisfies to basic assumptions of linear regression: the coefficients of variables are significant ($p$-value $<$ 0.01); the equation of regression is significant ($p$-value($F$-statistic) $<$ 0.01); the coefficients of variables have the right sign according to the economic substance; model residuals are normally distributed ($p$-value(JB) $>$ 0.05) and do not have strong autocorrelation (1.5 $\lesssim$ Durbin-Watson $\lesssim$ 2.5); the adjusted coefficient of determination is significant ($R_\mathrm{adj}^2>0.6$) and it does not differ sufficiently from $R^2$.

Also, factors included in the model do not have multicollinearity because the value of the variance inflation factor (VIF) is less than 10 for each factor \citep[Table~\ref{tab:tab4}, ][]{Shrestha2020}.

Figure~\ref{fig:fig13} shows the distribution of the model residuals. This distribution is approximately normal, satisfying the assumptions of linear regression. Additionally, we tested the normality of residuals using the library “SciPy”. This test compares calculated $p$-value with the significance threshold $\alpha$ (for this paper we chose $\alpha = 0.05$). If calculated $p$-value is more than $\alpha$ the hypothesis of normality of the distribution of model residuals is accepted (in our case $p$-value = 0.068)~\citep{AGOSTINO1971, Agostino1973}.

\begin{figure}[t]
    \includegraphics[width=\linewidth]{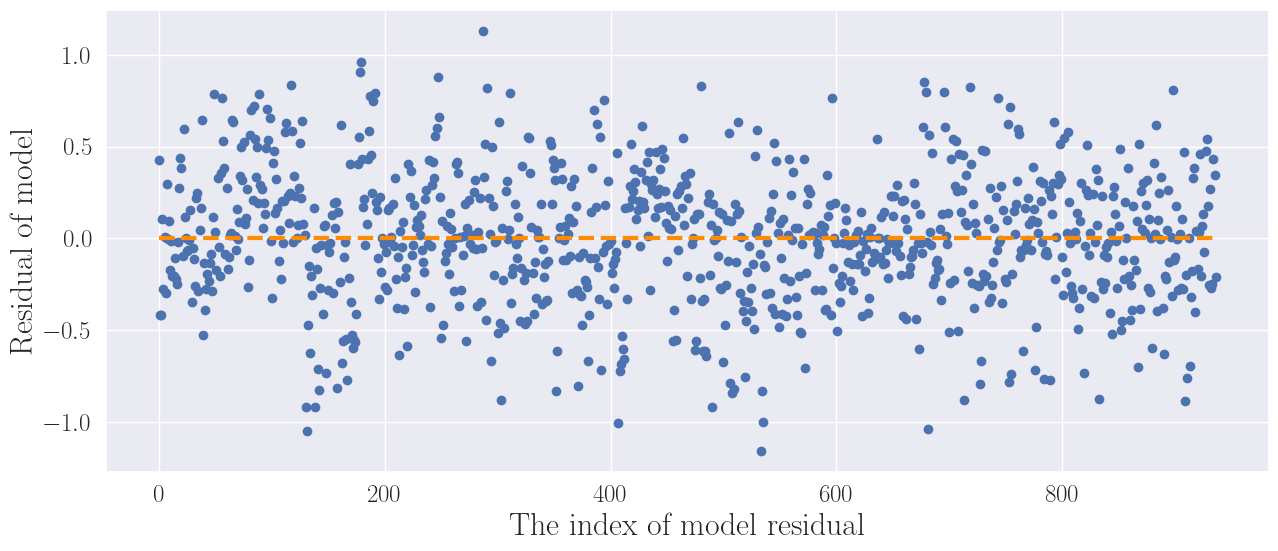}
    \caption{Homoscedasticity of the linear model residuals (segment “Land parcels”). The target variable is the natural logarithm of the per square meter price.}
    \label{fig:fig14}
\end{figure}

Also, we created the scatter plot of values of residuals against their numerical order. From Figure~\ref{fig:fig14} one can see that the data does not contain any dependencies. It means that the residuals are homoscedastic, and the last assumption of the linear regression is satisfied. 

On the base of the described results, we can state that our model is effective and interpretable. As we argued above, the key factor of the price formation for a land parcel is its location. Although the linear regression does relatively well, the price of an object can include latent (local-territorial) characteristics. In order to take this into account, we apply a powerful instrument dealing with spatially distributed data -- geostatistics. Geostatistical methods allow to model the distribution of objects in space. The central task of geostatistics is to reconstruct a studied phenomenon on the base of the values measured in a limited number of points \citep{Demyanov2010}. Mathematically, this problem can be contemplated as an interpolation problem. To get the best spatial estimations, the kriging method is used. Estimations obtained by the kriging have a minimal variation of error. Thus, the kriging is the best estimator in a class of linear interpolators \citep{Demyanov2010}. There are two basic conditions of usage of these methods -- the data should satisfy the spatial continuity and the stationarity conditions \citep{Kovalevskiy2012}.
The first condition means that shorter distances between the points should in general lead to closer values of the variable in these points.
The second condition means that in some locality the studied phenomenon must be spatially homogeneous.

\begin{figure}
    \centering
    \includegraphics[width=0.7\linewidth]{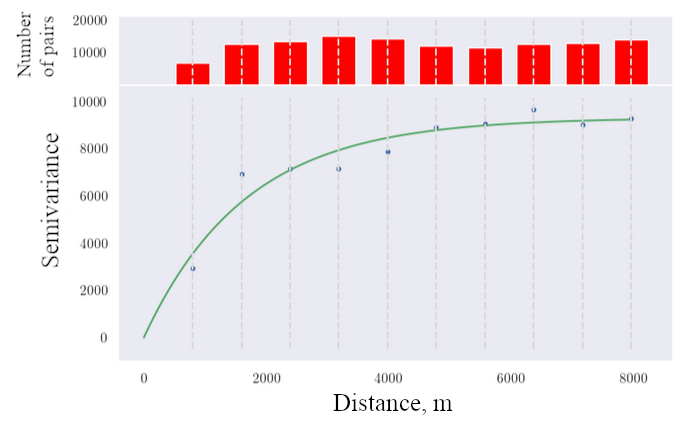}
    \caption{Variogram for modeling of the linear regression residuals by kriging method. The curve shows the fit of the points by an exponential function.}
    \label{fig:fig15}
\end{figure}

We check the condition of stationarity using the PDF of the distribution of the model residuals. If the histogram is close to the normal distribution, then the stationarity (statistical homogeneity) is assumed. Strictly speaking, this criteria is not sufficient for confirmation of stationarity, but it is strong enough for practical purposes \citep{Kovalevskiy2012}. Figure~\ref{fig:fig13} and the result of the Jarque–Bera test confirm the stationarity in the linear model residuals. It is worth noting that before checking the data on stationarity we need to examine whether the data contains a trend. In our case, we removed the trend from the data using linear regression, and then modeled the residuals with kriging. The trend in data can distort the variogram –  an instrument for estimation of spatial correlations in data~\citep{Demyanov2010,Isaaks1989}. The variogram model is a function describing dependence between the studied spatial variable and the distance. If data contains a trend, then the variogram would reproduce a large-scale trend that leads to missing of correlation of the observed variable on a small scale.

Another important (but also not sufficient) criteria of stationarity is that the variogram should have a sill beginning  at some distance. To build the variogram, the fitting curve should be chosen from a list of basic models. In this paper, we use the exponential model (see Figure~\ref{fig:fig15}).

The variogram is used to find the estimations of kriging \citep{Demyanov2010,Isaaks1989}, and it is substituted instead of the covariation function in the process of minimizing of the Lagrangian, which allows to obtain the interpolation coefficients of kriging taking into account constraints imposed on them \citep{Demyanov2010}.

\begin{figure}
    \includegraphics[width=\linewidth]{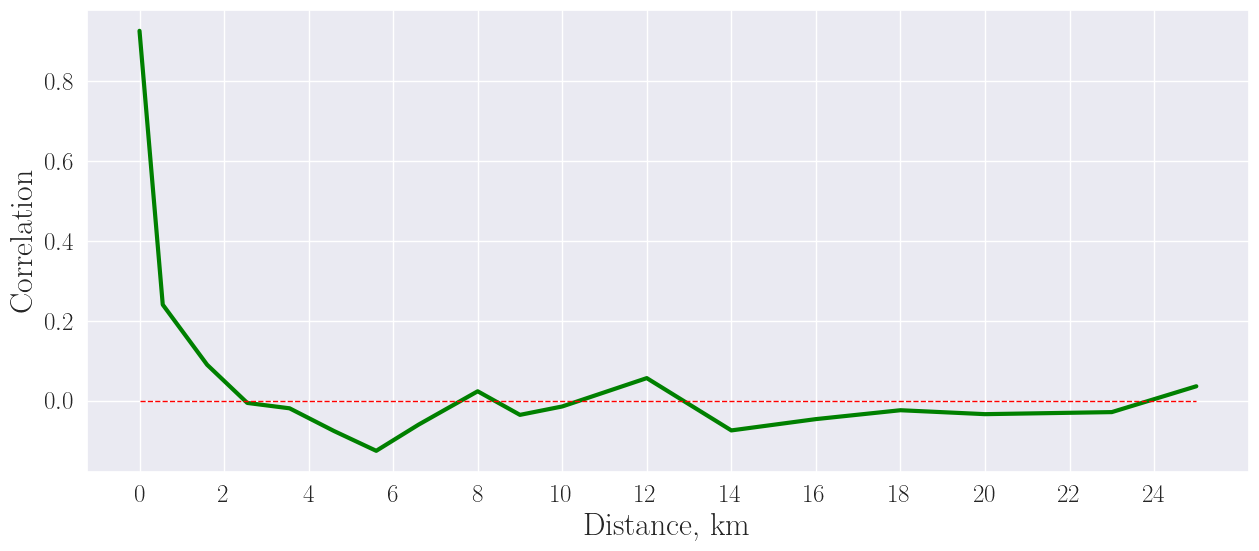}
    \caption{Check of the spatial continuity of the data. The correlation monotonically drops to zero and then fluctuates around this value.}
    \label{fig:fig16}
\end{figure}

The second condition for implementing kriging is the spatial continuity. We check it by plotting the correlation of points as a function of the distance between them. Figure~\ref{fig:fig16} shows that this condition is also satisfied.

Thus, the final (logarithmic) value of the per square meter price is the sum of the trend modeled by the linear regression and the residuals modeled with kriging. Then we exponentiate this value to obtain the final result.

\begin{figure*}[t]
    \centering
\begin{tabular}{cc}
	        \includegraphics[width=0.48\textwidth]{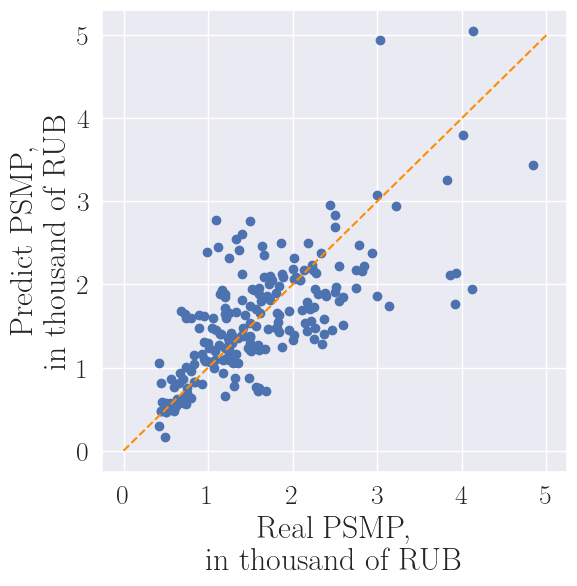}
	    &
	        \includegraphics[width=0.48\textwidth]{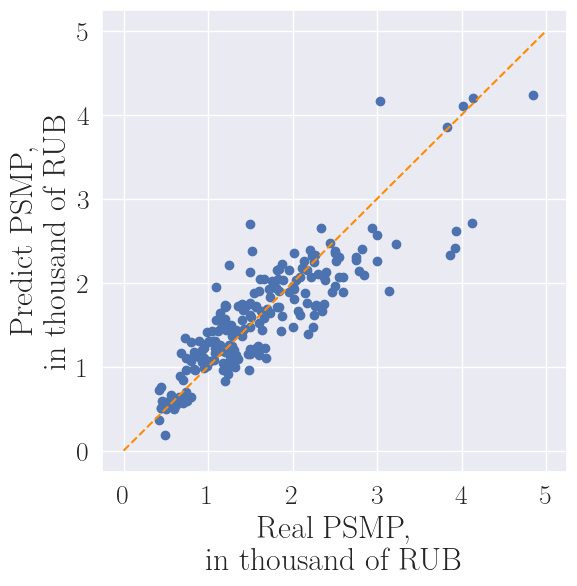}
	     \\
	        (a)
	    &
	        (b)
\end{tabular}
	\caption{Comparison between the model-predicted and the real values of the per square meter price (segment “Land parcels”). \textbf{(a)}: the linear regression, \textbf{(b)}: the linear regression with kriging interpolation of residuals.}
	\label{fig:fig17}
\end{figure*}

We assess this result by comparison of the fitting errors using cross-validation. The average fitting error is 28,36\% for linear regression and 19,23\% for regression-kriging. Figure~\ref{fig:fig17} shows the predicted values plotted against the real values. Both clouds of points are arranged along the diagonal line, but in the case of regression-kriging it has more elongated shape corresponding to lower errors.

\subsection*{segment “Flats”}

\begin{table}[!t]
    \caption{Results of training of the linear regression (segment “Flats”).}
    \label{tab:tab5}
    \begin{tabular}{|>{\centering\arraybackslash}m{3cm}>{\centering\arraybackslash}m{1.95cm}>{\centering\arraybackslash}m{1.95cm}>{\centering\arraybackslash}m{1.95cm}>{\centering\arraybackslash}m{1.95cm}|}
     \hline
     \multicolumn{5}{|c|}{\textbf{OLS Regression Results}} \\
     \hline
     \multicolumn{1}{|c}{Dep. variable:} & \multicolumn{4}{c|}{$y$} \\
     \multicolumn{5}{|c|}{} \\
     \multicolumn{1}{|c}{$R^2$:} & \multicolumn{4}{c|}{0.319} \\
     \multicolumn{1}{|c}{$R_{adj}^2$:} & \multicolumn{4}{c|}{0.318} \\
     \multicolumn{1}{|c}{$F$-statistic:} & \multicolumn{4}{c|}{227.8} \\
     \multicolumn{1}{|c}{$p$-value ($F$-statistic):} & \multicolumn{4}{c|}{$\ll$ 0.001} \\
     \hline
      & \centering\textit{coef} & \centering\textit{std err} & \centering\textit{t} & $P>|t|$ \\
     \hline
     \centering{$const$} & \centering{1.538$e$+05} & \centering{378.085} & \centering{406.867} & $\ll 0.001$ \\
     \centering{$x_1$} & \centering{-8730.2980} & \centering{384.482} & \centering{-22.707} & $\ll 0.001$ \\
     \centering{$x_3$} & \centering{3472.7788} & \centering{552.203} & \centering{6.289} & $\ll 0.001$ \\
     \centering{$x_4$} & \centering{-1075.3009} & \centering{465.506} & \centering{-2.310} & $0.021$ \\
     \centering{$x_7$} & \centering{-3200.5609} & \centering{445.592} & \centering{-7.183} & $\ll 0.001$ \\
     \centering{$x_8$} & \centering{3374.1161} & \centering{495.161} & \centering{6.814} & $\ll 0.001$ \\
     \centering{$x_{10}$} & \centering{4874.6153} & \centering{439.653} & \centering{11.087} & $\ll 0.001$ \\
     \centering{$x_{11}$} & \centering{-4124.0290} & \centering{455.890} & \centering{-9.046} & $\ll 0.001$ \\
     \centering{$x_{18}$} & \centering{-5892.8762} & \centering{384.745} & \centering{-15.316} & $\ll 0.001$ \\
     \multicolumn{5}{|l|}{$y$ is the Per square meter price} \\ 
     \multicolumn{5}{|l|}{$x_1$ is Area of flat} \\ 
     \multicolumn{5}{|l|}{$x_3$ is Distance to the water objects (straight-line distance)} \\ 
     \multicolumn{5}{|l|}{$x_4$ is Distance to the seacoast (straight-line distance)} \\ 
     \multicolumn{5}{|l|}{$x_7$ is Distance to the closest railway station (straight-line distance)} \\ 
     \multicolumn{5}{|l|}{$x_8$ is Distance to the area of negatively influencing factors} \\
     \multicolumn{5}{|l|}{\phantom{lllll}(straight-line distance)} \\
     \multicolumn{5}{|l|}{$x_{10}$ is Number of  cafes (within a radius of 1 km)} \\ 
     \multicolumn{5}{|l|}{$x_{11}$ is Distance to the town hall (distance by the roads)} \\ 
     \multicolumn{5}{|l|}{$x_{18}$ is Wall material: large-panel} \\
     \hline
     \multicolumn{1}{|c}{Durbin-Watson:} & \multicolumn{4}{c|}{1.952} \\
     \multicolumn{1}{|c}{Jarque-Bera (JB):} & \multicolumn{4}{c|}{791.269} \\
     \multicolumn{1}{|c}{$p$-value (JB):} & \multicolumn{4}{c|}{$\ll$ 0.001} \\
     \multicolumn{1}{|c}{Cond. No.:} & \multicolumn{4}{c|}{2.96} \\
     \hline
    \end{tabular}
\end{table}

As it is described in Section~\ref{select_features}, the correlations between the dependent variable and the pricing variables for flats are significantly smaller than for land parcels. There is no explicit dependence on the object location. Nevertheless, socially important factors such as “Number of shopping malls (within a radius of 1 km)” and “Number of cafes (within a radius of 1 km)” have higher correlations with the per square meter price than other factors, although still insufficient. However, we start with an effort to create a simple and interpretable model –- the linear regression. This step is necessary according to the methodological rules of mass valuation of real estate~\citep{Method2018}. Before modeling, the data was prepared by selecting the factors (features) included in the model and performing the data standardization~\citep{Konayeva2024}. The standardization is useful in the case of considerably different values of factors in order to reduce them to one scale. 
\begin{figure}[t]
    \includegraphics[width=\linewidth]{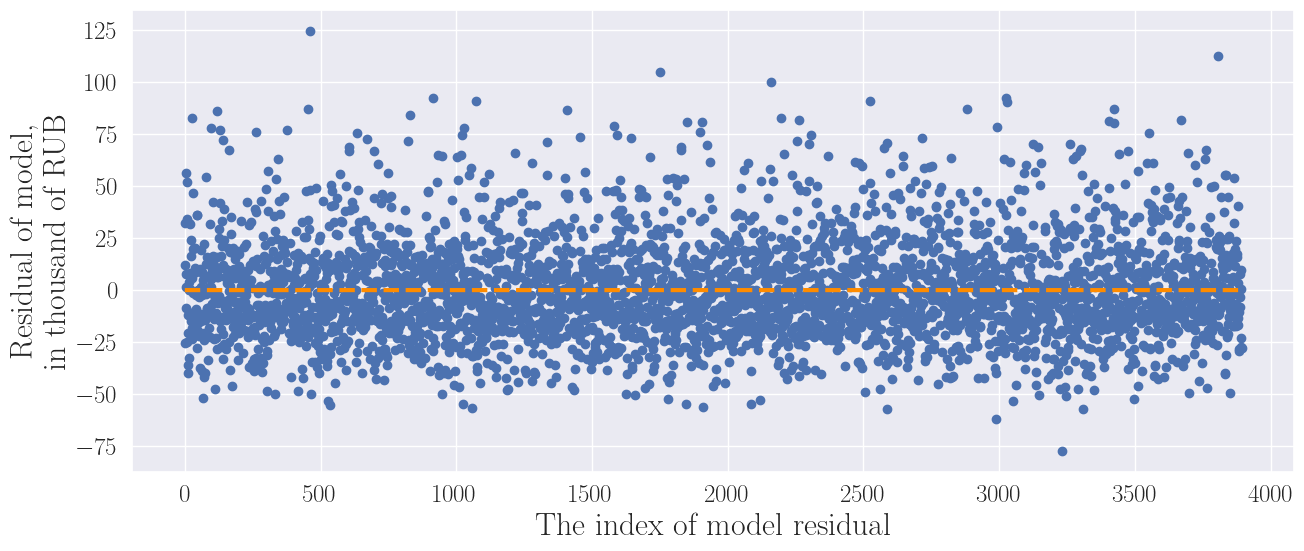}
    \caption{Homoscedasticity of the linear model residuals (segment “Flats”).}
    \label{fig:fig18}
\end{figure}

\begin{figure}
    \includegraphics[width=\linewidth]{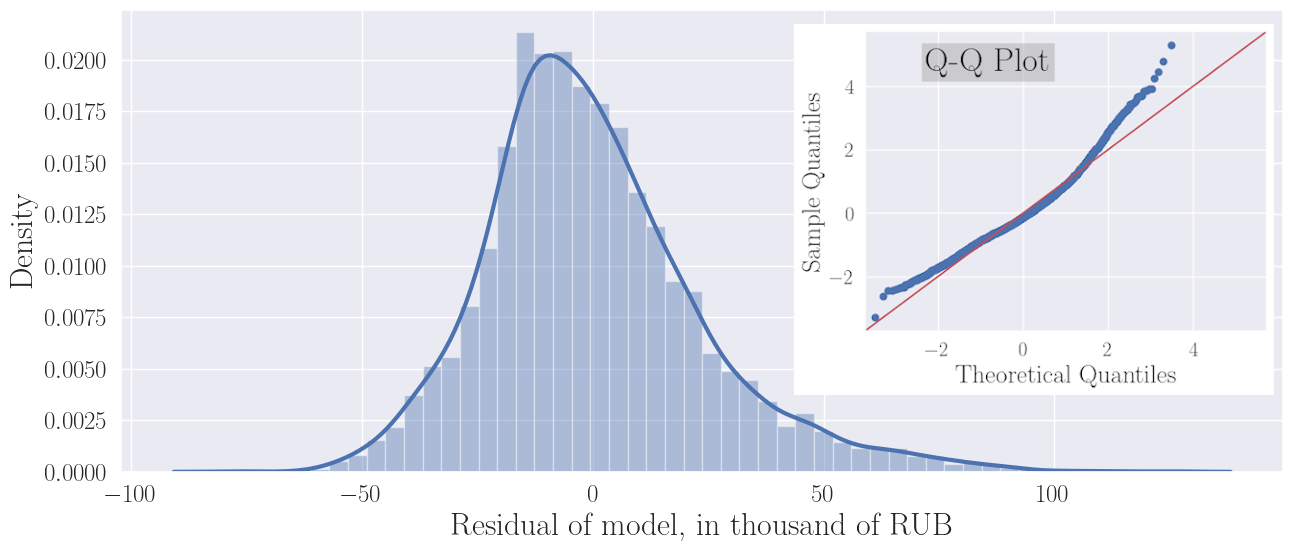}
    \caption{Distribution of the linear model residuals (segment “Flats”). The inset shows Quantile-Quantile plot. The tails of the sample distribution are clearly inconsistent with the normal distribution.}
    \label{fig:fig19}
\end{figure}

The results of the model with eight most significant features selected with SelectKBest algorithm are shown in Table~\ref{tab:tab5}.

From Table~\ref{tab:tab5} we can see that the coefficients of variables are significant ($p$-value $<$ 0.05); the regression is significant ($p$-value($F$-statistic) $<$ 0.01); the coefficients of variables have the right sign according to economic substance; model residuals  do not have autocorrelation (1.5 $\lesssim$ Durbin-Watson $\lesssim$ 2.5); the value of Condition Number~\citep{Gujarati2009} is 2.96, it means that factors do not contain multicollinearity.
Also, we checked that the expected value of residuals is close to zero and the dispersion of residuals is constant (Figure~\ref{fig:fig18}).

However, the residuals of the model are not normally distributed ($p$-value (JB) $<$ 0.05, see also Figure~\ref{fig:fig19}). It means that the estimates obtained with the OLS are not optimal -- they do not have minimal dispersion in the class of linear unbiased estimators. Also, the explanatory power of the model is poor since the value of $R_{adj}^2=0.318$ is low.

\begin{figure*}[t]
    \centering
\begin{tabular}{cc}
	        \includegraphics[width=0.48\textwidth]{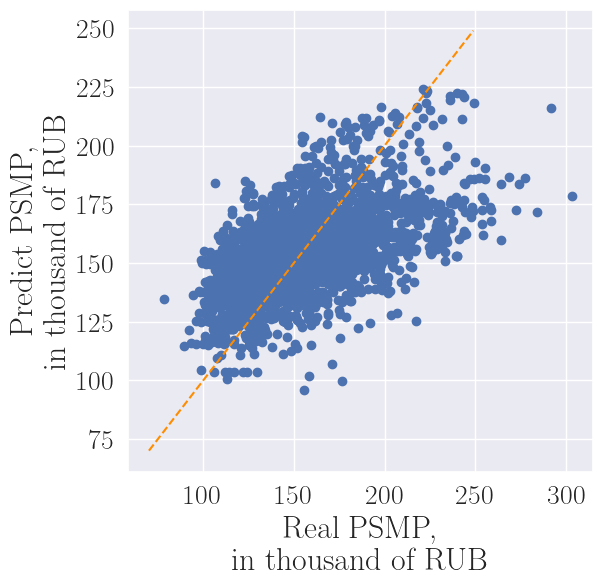}
	    &
	        \includegraphics[width=0.48\textwidth]{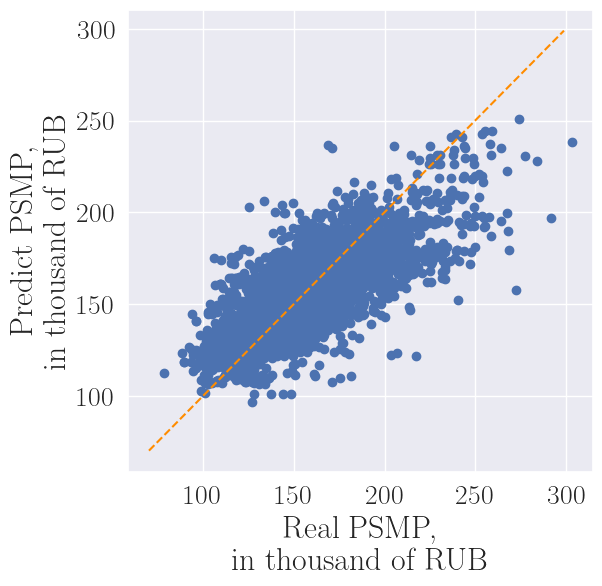}
	     \\
	        (a)
	    &
	        (b)
\end{tabular}
	\caption{Comparison between the model-predicted and the real values of the per square meter price (segment “Flats”). \textbf{(a)}: linear regression, \textbf{(b)}: RuleFit.}
	\label{fig:fig20}
\end{figure*}

Figure~\ref{fig:fig20} (a) shows comparison between the predicted and the real values of the per square meter price.
The model clearly overestimates lower values and underestimates higher ones.

We summarize the results of the linear regression model in Table~\ref{tab:tab6}. Later they will be compared with other models through the following metrics: $R^2_\mathrm{adj}$, mean absolute error (MAE)

\begin{equation}
   \text{MAE}=\frac{1}{N}\sum_{i=1}^{N}|e_{i}|=\frac{1}{N}\sum_{i=1}^{N}|y_i-\hat{y_i}|,
   \label{eq:mae}
\end{equation}

and mean absolute percentage error (MAPE)

\begin{equation}
   \text{MAPE}=\frac{1}{N}\sum_{i=1}^{N}\frac{|y_i-\hat{y_i}|}{y_i}\times100\%.
   \label{eq:avg_err}
\end{equation}

\begin{table}[h!]
  \begin{center}
    \caption{Basic results of the linear regression (segment “Flats”).}
    \label{tab:tab6}
    \begin{tabular}{|>{\centering\arraybackslash}m{0.14\linewidth}|>{\centering\arraybackslash}m{0.13\linewidth}|>{\centering\arraybackslash}m{0.17\linewidth}|>{\centering\arraybackslash}m{0.15\linewidth}|>{\centering\arraybackslash}m{0.11\linewidth}|>{\centering\arraybackslash}m{0.11\linewidth}|} 
    \hline
      \rowcolor{lightgray}
      \textbf{$R_\mathrm{adj}^2$ train dataset} & \textbf{$R_\mathrm{adj}^2$ test dataset} & \textbf{MAE train dataset} & \textbf{MAE test dataset} & \textbf{MAPE train dataset} & \textbf{MAPE test dataset} \\
      \hline
      0.319 & 0.339 & 17 890.32 & 17 947.46 & 11.7\% & 11.9\%\\
      \hline
    \end{tabular}
  \end{center}
\end{table}

The linear regression becomes ineffective in the presence of non-linear dependencies between variables. There are more suitable algorithms, among which the most simple is the ``Decision tree'' \citep{DiazRamirez2023}. This algorithm forms the hierarchic tree structure that consists of ``If … then …'' rules. The rules are generated in the process of training on the base of the training dataset. They are created in natural language and simple for understanding. This is an advantage of the algorithm. But there is a significant drawback -- the algorithm is prone to overfitting, meaning that the generated rules only fit the training dataset and show poor results on others. The extension of this algorithm called “Random Forest”~\citep{Breiman2017} rectifies this shortcoming. This method creates a set of decision trees. The advantage of it is high accuracy of prediction and good performance under default hyperparameters. Compared to the “Decision tree”, the “Random Forest” is much less prone to overfitting. The limitation of this method is complicated interpretation of the modeling results due to a large number of generated rules.

Yet one algorithm that predicts with good accuracy and possesses better interpretability than “Random Forest” is “RuleFit” ~\citep{Molnar2022,Masis2021}. The underlying concept of the method is a combination of linear regression and prediction models based on decision trees. This model was developed by \cite{Friedman2008} as a sparse linear regression model (a class of models allowed to have coefficients with zero weights after training). This algorithm detects the effects of interaction between features in the form of decision rules automatically generated from decision trees. The set of decision trees can be generated by any appropriate algorithm, e.g.  the Random Forest or the Gradient Boosting. The resulting rules are used in the linear regression model as individual variables. The base model of the algorithm is the LASSO \citep[][see Eq.~\ref{eq_lasso}]{Tibshirani1996}, because it allows to include only the most significant rules in the final model. The goal is to reach a compromise between the errors of regression and the dimension of the feature space. Therefore, the RuleFit is the LASSO regression that takes as an input both the individual factors and the rules created by the Random Forest and determines the most informative among them. The tendency of some factors to disappear is controlled by a regularization parameter that is fitted using cross-validation \citep[more information can be found in ][]{{Friedman2010}}.
Unlike the OLS, the LASSO (and consequently the RuleFit) does not have any prerequisites. The effectiveness of the model can be estimated by common metrics such as $R^2_\mathrm{adj}$, MAE, and MAPE.

To build the RuleFit model, we used both binary and continuous features. As continuous features, we used the factors from the correlation matrix (Figure~\ref{fig:fig10}) having the correlation with the dependent variable 0.15 or more. The following factors satisfy this condition:

\begin{description}
 \setlength\itemsep{-0.3em}
 \item $x_1$: Area of flat 
 \item $x_2$: Distance to the hospitals (straight-line distance) 
 \item $x_3$: Distance to the water objects (straight-line distance) 
 \item $x_4$: Distance to the seacoast (straight-line distance) 
 \item $x_6$: Distance to the business activities center (straight-line distance) 
 \item $x_7$: Distance to the closest railway station (straight-line distance) 
 \item $x_8$: Distance to the area of negatively influencing factors (straight-line distance) 
 \item $x_9$: Number of shopping malls (within a radius of 1 km) 
 \item $x_{10}$: Number of  cafes (within a radius of 1 km) 
 \item $x_{11}$: Distance to the town hall (distance by the roads) 
 \item $x_{15}$: Aggregate: Year built \& Number of storeys 
\end{description}

As binary  features we used the factors describing the storey of the flat (“the first”, “the last”) and the wall material of the building (“bricky”, “large-panel”, “monolithic”, “small-element concrete blocks”). Thus, we used 17 features for modeling. We created all kinds of combinations (each with the number of factors $<$ 12, due to computational limitations) of these factors using the RuleFit. To select the best model, we used the values of $R^2_\mathrm{adj}$ and MAE, and also checked for concordance of the variable signs with their economic substance. In order to keep the interpretability of the model, we limited the number of rules to 50. For generation of rules, the Random Forest algorithm was applied. The final model contains 22 rules and 12  features. They are shown in Table~\ref{tab:tab9} of ~\ref{rf_res}.

The model contains both the factors describing the basic characteristics of a building, which allowed to differentiate one building from another, and factors characterizing the location of a building. It also distinguishes individual characteristics of a flat within one building. The majority of rules include the factors “Distance to the town hall”, “Aggregate: Year built \& Number of storeys”, and “Area of flat”. It can be explained by the fact that Vladivostok is a highly diverse city with entwinement of historical and modern buildings, and the RuleFit tries to differentiate between them. This is a typical example of a situation when the linear regression performs poorly, and grasping of non-linear dependencies is required.

Figure~\ref{fig:fig20} (b) shows the comparison between the predicted and the real values of the per square meter price. The RuleFit clearly demonstrates better performance compared to the OLS regression in terms of the residual variance. It still underestimates higher prices, but shows decent results for lower ones.

Like any other model, the RuleFit has its drawbacks. First, it sometimes keeps rules that are very similar to each other (see Table~\ref{tab:tab9}), which complicates the interpretation. Second, due to the stochastic nature of the tree generation process, the RuleFit every time generates new rules (this problem can be solved by explicitly setting the state of the random number generator). Additionally, each training dataset leads to a different set of final rules. Thus, the RuleFit is a badly reproducible model.

\begin{table}[t]
  \begin{center}
    \caption{Basic results of the RuleFit model (segment “Flats”).}
    \label{tab:tab7}
    \begin{tabular}{|c|c|c|} 
    \hline
      \rowcolor{lightgray}
      \textbf{$R_\mathrm{adj}^2$} & MAE & MAPE\\
      \hline
     0.6 & 13 518.28 & 8.8\% \\
      \hline
    \end{tabular}
  \end{center}
\end{table}

We present the main results in Table~\ref{tab:tab7}. The coefficient of determination is equal to 0.6, which is acceptable for our purposes. The values of MAE and MAPE for the  RuleFit are lower than for the linear regression. 
Therefore, the RuleFit seems to be a perspective method for prediction of the per square meter price. Despite its drawbacks mentioned above, the RuleFit model fairly well describes the data while maintaining the interpretability.

\section{Discussion and conclusions}
\label{conc}

In this paper we overview modern approaches to building interpretable models of real estate market, discuss their pros and cons in the context of cadastral valuation, describe all stages of modeling process with special focus on data preprocessing and feature engineering, and build inherently interpretable models for two large and very different market segments: the non-commercial land parcels and flats in residential buildings. The models presented in this work were used for cadastral valuation of 700 000 flats and more than 570 000 land parcels over the whole territory of the Primorsky Krai (Russia), highlighting the practical value of the research for public and legal policies. Our results encourage usage of hybrid approaches, which for both market segments show better performance than traditional single component models. We believe that our approach can be applied for property markets in other geographical areas without significant changes, and encourage further research.

\begin{table}[t]
  \begin{center}
    \caption{Basic results of the Random Forest algorithm.}
    \label{tab:tab8}
    \begin{tabular}{|c|c|c|c|} 
    \hline
      \rowcolor{lightgray}
      \textbf{Segment} & {$R_\mathrm{adj}^2$} & MAE & MAPE\\
      \hline
        Land parcels & 0.75 & 586.69 & 20.36\% \\
      \hline
        Flats & 0.73 & 10830.45 & 7.01\% \\
      \hline
    \end{tabular}
  \end{center}
\end{table}

\begin{figure*}[t]
    \centering
\begin{tabular}{cc}
	        \includegraphics[width=0.48\textwidth]{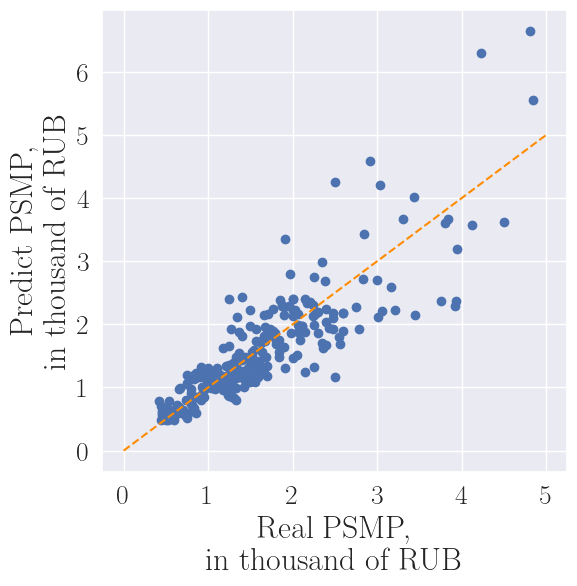}
	    &
	        \includegraphics[width=0.5\textwidth]{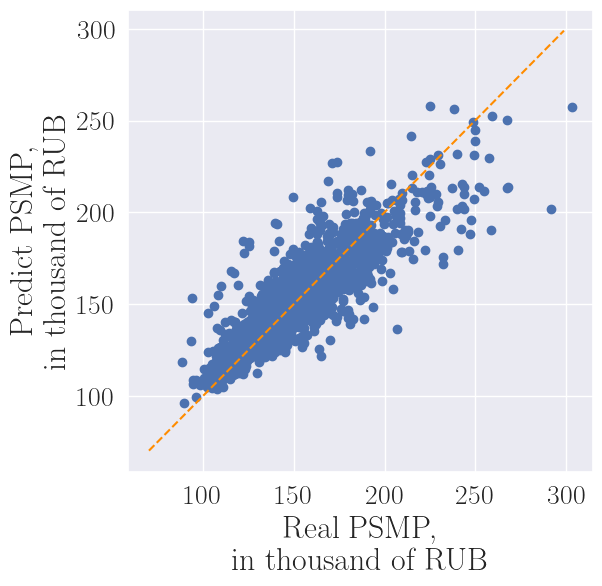}
	     \\
	        (a)
	    &
	        (b)
\end{tabular}
	\caption{Results of modeling using Random Forest algorithm. \textbf{(a)}: the segment ``Land Parcels'', \textbf{(b)}: the segment ``Flats''.}
	\label{fig:fig21}
\end{figure*}

An important aspect of the presented analysis is comparison of performance of the proposed intrinsically interpretable models with the ``black-box'' models renowned for their ability to catch complex patterns in the data. As an example of such models, we consider one of the most popular algorithms~-- the Random Forest, and build models for both market segments using default hyperparameters, which is enough to obtain rough estimations. The results are summarized in Table~\ref{tab:tab8}. For land parcels, the Random Forest unsurprisingly outperforms the OLS regression with MAPE of 20.36\% vs 28.36\% and similar $R^2$. However, its MAPE is approximately equal to the corresponding error of the regression-kriging (19.23\%) with potential for minor improvement after hyperparameter tuning. These numbers demonstrate that the regression-kriging is approximately equivalent to the Random Forest in terms of precision, but has better interpretability as being based on the linear regression. The scatter plots for the regression-kriging (Figure~\ref{fig:fig17}, b) and the Random Forest (Figure~\ref{fig:fig21}, a) are also very similar. For flats, the situation is not so clear. The Random Forest has the MAPE of 7.0\% vs 8.8\% for the RuleFit. Also, the $R^2$ for the former is higher: 0.73 vs 0.60. Additionally, visual inspection of the scatter plots shows that the points for the Random Forest (Figure~\ref{fig:fig21}, b) group more tightly (and approximately symmetrically) along the diagonal line comparing to the RuleFit (Figure~\ref{fig:fig20}, b), which tends to overestimate lower values and underestimate higher ones. While the Random Forest shows slightly better results, the RuleFit has an important advantage of ad hoc interpretability. The final choice between these approaches depends on whether or not the post-hoc nature of the Random Forest SHAP interpretability is acceptable for a certain task. It gives rise to a very important question outlined below.

With fast development of machine learning, an increasing  number of concerns emerges about legal aspects and practical implementations of machine learning research. However, the main focus of these discussions is on the ethical sides such as privacy and responsible data collection and storage \citep{Choy2023}. While we do not intend to diminish the importance of this problem, we encourage to develop a legal base for integration of machine learning to state policies like the cadastral valuation. This is a highly inert area, which still broadly uses classical and, to a large extent, obsolete methods of real estate valuation. Such a situation hinders further development of the field and leads to inferior results of cadastral valuation and, eventually, non-optimal taxes, since modern and efficient ML methods are not protected by law. As it was mentioned previously, in the case of litigation, the court prefers well documented traditional arguments. This is the law policy in Russia, which may differ for some other countries with policies better meeting modern requirements. We expect further research and discussions on this topic.

Apart from considering various models and their implementations, the study addresses an important topic of data preparation. In practice, real data often suffer from heavy noise due to insufficiently developed policies for data collection. Standard treatment of outliers is not enough for such cases, so an advanced method is suggested. It consists of preparatory clusterization of points both by their location (to deal with spatial heterogeneity) and by their characteristics (to form clusters of similar objects). Then the iterative search for outliers is implemented by training of the linear regression model robust to outliers (RANSAC regression), specially designed to deal with problematic datasets. While it allowed to significantly reduce the price discrepancy between similar objects, further systematic study is needed. Another problem addressed in our study is rigorous formulation of the concept of traffic convenience referred here as ``Development of road network''. Surprisingly, this topic is not sufficiently covered in literature despite its potential influence on price formation of property objects. Suggested in this paper simple intuitive formulation through the concept of centrality from graph theory shows weak correlation of this feature with the target variable for Vladivostok. It was not included in the final model for this city due to the presence of features with higher correlation, but it was used to build models for two other cities of Primorsky Krai, Artyom and Nakhodka (see Figure~\ref{fig:about_prymorye}). The concept requires further research to better reflect the real situation and, possibly, capture not only the development of road network, but also public transport availability. Finally, we note that our study could benefit from adding to consideration the temporal price dynamics. The current research relies on the one-year data due to the absence of structured data for previous periods, but potentially incorporating such data into the modeling process can make predictions more precise.

\section*{Acknowledgments}
\label{acknldg}
We thank Artyom S. Tanashkin for thorough reading of the preprint and providing numerous suggestions, which improved the readability and clarity of the manuscript. We are also grateful to the anonymous reviewers for their interest and invaluable feedback, which allowed to significantly improve the scientific value of the paper.

Maps are drawn using OpenStreetMap and Python library Folium. For data analysis and modeling we used the following Python libraries (listed in the alphabet order): GeoPy, Matplotlib \citep{Hunter:2007}, NetworkX, NumPy \citep{harris2020array}, OSMnx, pandas \citep{mckinney-proc-scipy-2010}, RuleFit~\citep{rulefit}, SciKit-learn, SciKit GStat~\citep{Maelicke2021}, SciPy, seaborn \citep{Waskom2021}, statsmodels.

\bibliographystyle{elsarticle-harv}
\bibliography{paper.bib}

\appendix
\section{The resulting model for RuleFit}
\label{rf_res}

The final model for RuleFit is given by:

\begin{equation}
   \hat{y} = 187490.76 + \sum_{i=1}^{34}\alpha_if_i
   \label{eq:RF}
\end{equation}

where $\alpha_i$ are coefficients assigned by the RuleFit; $f_i$ - factor or rule from Table~\ref{tab:tab9} and the naming of factors is given below.

\begin{description}
 \setlength\itemsep{-0.3em}
 \item $x_1$: Area of flat 
 \item $x_3$: Distance to the water objects (straight-line distance) 
 \item $x_4$: Distance to the seacoast (straight-line distance) 
 \item $x_6$: Distance to the closest business activities center (straight-line distance) 
 \item $x_7$: Distance to the closest railway station (straight-line distance) 
 \item $x_{10}$: Number of  cafes (within a radius of 1 km) 
 \item $x_{11}$: Distance to the town hall (distance by the roads) 
 \item $x_{15}$: Aggregate: Year built \& Number of storeys
 \item $x_{19}$: Wall material: bricky
 \item $x_{20}$: Wall material: monolithic
 \item $x_{21}$: Storey of flat: the first
 \item $x_{22}$: Storey of flat: the last
\end{description}

\begin{table}
  \begin{center}
    \caption{Rules and features for the final RuleFit model (segment “Flats”).}
    \label{tab:tab9}
    \begin{tabular}{|c|c|c|c|} 
    \hline
      \rowcolor{lightgray}
      \textbf{№} & \textbf{Factor/Rule $f$} & \textbf{Type} & \textbf{Coefficient $\alpha$}  \\
      \hline
      1 & $x_{15}$ & linear & 6713.23 \\
      2 & $x_1$ & linear & -438.15 \\
      3 & $x_4$ & linear & -3.14 \\
      4 & $x_{19}$ & linear & 6482.54 \\
      5 & $x_7$ & linear & -2.83 \\
      6 & $x_{20}$ & linear & 7435.15 \\
      7 & $x_3$ & linear & 2.02 \\
      8 & $x_{10}$ & linear & 76.72 \\
      9 & $x_{11}$ & linear & -0.24 \\
      10 & $x_{21}$ & linear & -2771.67 \\
      11 & $x_{22}$ & linear & -387.96 \\
      12 & $x_6$ & linear & -0.08 \\
      13 & $x_{11} > 3710.14 \text{ \& } x_1 > 41.75 \text{ \& } x_6 > 203.57 \text{ \& } x_{15} > 1.43$ & rule & -26328.13 \\
      14 & $x_{15} > 1.95 \text{ \& } x_{11} \leq 3710.14$ & rule & 29815.42 \\
      15 & $x_{11} > 3699.71 \text{ \& } x_1 > 40.85$ & rule & -6277.76 \\
      16 & $x_1 \leq 39.8 \text{ \& } x_{15} \leq 1.95 \text{ \& } x_{11} \leq 3710.14 \text{ \& } x_4 \leq 641.88$ & rule & 17615.75 \\
      17 & $x_{11} \leq 3706.48 \text{ \& } x_{10} > 40.5$ & rule & 13697.61 \\
      18 & $x_{11} > 3710.14$ & rule & -6846.8 \\
      19 & $x_1 \leq 41.75 \text{ \& } x_{11} > 3710.14 \text{ \& } x_{15} \leq 1.11$ & rule & 8955.08 \\
      20 & $x_1 \leq 41.75 \text{ \& } x_{11} > 3710.14 \text{ \& } x_1 > 18.84 \text{ \& } x_{15} \leq 1.11$ & rule & -4897.04 \\
      21 & $x_{15} \leq 0.59 \text{ \& } x_{11} > 3741.63 \text{ \& } x_1 > 42.14$ & rule & -4758.11 \\
      22 & $x_{11} > 3710.14 \text{ \& } x_1 > 41.64 \text{ \& } x_{15} > 1.43$ & rule & -7064.57 \\
      23 & $x_1 > 41.75 \text{ \& } x_{11} > 3710.14 \text{ \& } x_{15} \leq 1.43$ & rule & -3785.21 \\
      24 & $x_{15} > 0.59 \text{ \& } x_{11} > 3741.63 \text{ \& } x_1 > 42.14$ & rule & 4401.15 \\
      25 & $x_1 > 40.95 \text{ \& } x_{11} \leq 6492.79 \text{ \& } x_{11} > 3706.48 \text{ \& } x_{15} > 0.59$ & rule & 6457.31 \\
      26 & $x_{10} \leq 32.5 \text{ \& } x_{15} \leq 1.95 \text{ \& } x_{11} \leq 3710.14 \text{ \& } x_1 > 39.80$ & rule & -3693.24 \\
      27 & $x_{11} \leq 3644.45 \text{ \& } x_3 \leq 2305.70$ & rule & 2686.18 \\
      28 & $x_{11} \leq 6492.79 \text{ \& } x_{11} > 3710.14 \text{ \& } x_1 > 41.64 \text{ \& } x_{15} > 1.43$ & rule & 5532.75 \\
      29 & $x_{11} \leq 6492.79 \text{ \& } x_{11} > 3710.14 \text{ \& } x_1 > 41.75 \text{ \& } x_{15} > 1.43$ & rule & 4444.61 \\
      30 & $x_{1} > 40.95 \text{ \& } x_{11} > 3706.48 \text{ \& } x_{15} \leq 0.59$ & rule & -898.74 \\
      31 & $x_{11} \leq 3637.04$ & rule & 718.14 \\
      32 & $x_1 > 40.95 \text{ \& } x_{11} > 3637.04$ & rule & -387.44 \\
      33 & $x_{11} \leq 3710.14$ & rule & 399.47 \\
      34 & $x_{15} > 1.95 \text{ \& } x_{11} \leq 3637.04$ & rule & 16.4 \\
      \hline
    \end{tabular}
  \end{center}
\end{table}

\end{document}